\documentclass[
 reprint,
 amsmath,amssymb,
 aps,
 prb
]{revtex4-2}

\usepackage{graphicx}
\usepackage{dcolumn}
\usepackage{bm}
\usepackage{color}
\usepackage[colorlinks=true,linkcolor=blue,allcolors=blue]{hyperref}

\begin{document}

\preprint{prb}

\title{Spectral properties of disordered insulating lattice under nonlinear electric field}

\author{Kunal Mozumdar}
\affiliation{%
 Department of Physics, University at Buffalo,
  SUNY, Buffalo NY 14260}
 \author{Herbert F. Fotso}%
\affiliation{%
 Department of Physics, University at Buffalo,
  SUNY, Buffalo NY 14260}
 
\author{Jong E. Han}%
 \email{jonghan@buffalo.edu}
\affiliation{%
 Department of Physics, University at Buffalo,
  SUNY, Buffalo NY 14260}

\date{\today}

\begin{abstract}
Quenched disorder in a solid state system can result in Anderson localization, where electrons are exponentially localized and the system behaves like an insulator. By solving exactly a disordered electronic lattice model out of equilibrium, we investigate the effect of a DC electric field on Anderson localization in an open system, and provide a minimal platform to study disorder-nonequilibrium interplay in electronic lattice systems. We perform steady-state Keldysh Green's function calculations on an infinite lattice with a finite-range of disorder-active region that are coupled to fermion reservoirs. Our solutions out of a fully electronic model verify Mott's temperature scaling of the variable-range-hopping transport and the Lifshitz tail, well-corroborated by the coherent-potential approximation. We further reveal that a nonequilibrium electronic lattice creates a statistical evolution that shows a counterintuitive shift of the distribution edge in the opposite direction of the band edge. The rich evolution of non-thermal statistics highlights the importance of an explicit band structure and the impurity correlations in strong nonequilibrium theories.
\end{abstract}

\maketitle


\section{\label{sec:level1}Introduction}
Disordered solid-state systems have been a problem of great interest in condensed matter physics. Seminal work by P. W. Anderson in 1958 \cite{anderson1958absence} showed that in a regular lattice with disordered potential, there is the absence of diffusion of the electronic wavefunctions, which get confined in certain regions of the lattice irrespective of the underlying distribution of disorder. The Anderson localization arises from the quantum interference of electronic wavefunctions mixing at random energy levels. This groundbreaking concept, primarily discussed in the context of electronic systems \cite{RevModPhys.57.287,mott1967electrons,mott1968conduction1, thouless1974electrons,kramer1993localization,anderson201050,cutler1969observation}, has since been extended to various wave phenomena \cite{sheng1990scattering}, including acoustic \cite{kirkpatrick1985localization}, electromagnetic \cite{john1984electromagnetic,lagendijk2009fifty,wiersma1997localization, schwartz2007transport,segev2013anderson}, gravitational waves\cite{rothstein2013gravitational}. It is relevant for applications in electronic devices \cite{tian2019anderson} and photonic materials~\cite{mafi2019disordered}, etc. Almost a decade after Anderson's paper, Neville Mott argued that Anderson localization is the mechanism of disorder driven metal to insulator transition called the Anderson Transition \cite{mott1975anderson,mott1971conduction, Mott01041961}, which happens over a mobility edge, the energy scale below which a particle is localized. Fluctuations in the random disordered potential allow localized levels to appear near the band-edge which form Lifshitz tails~\cite{halperin1965green, garcia2024fluctuation, RevModPhys.64.755} and the mobility edge separates these localized states from the delocalized extended states. 

A much less studied problem is the effect of a DC electric field on Anderson localization. In disordered materials, the electric field influences the phase coherence lengths that can affect Anderson localization~\cite{lee1983effects, soukoulis1983localization}. Some earlier analytic studies~\cite{prigodin1980one,kirkpatrick1986anderson} have reported that in a weak field, there is a power-law localization instead of Anderson localization. At some stronger critical field, there is a mobility edge beyond which the states are extended. Other approaches~\cite{prigodin1989localization} calculate the electron density fluctuation and relaxation dynamics showing delocalization in the presence of strong fields. In a weakly disordered two-dimensional electronic system, it was claimed that a very small electric field can disrupt localization \cite{bleibaum1995theory, bleibaum2004weak}. In this work, we advance the understanding of the nonlinear transport in electronic lattice. The goals of this work are (1) to establish a basic lattice model and its solutions as a framework to study disordered nonequilibrium electronic lattice, (2) to understand how an electric field delocalizes a disordered system through signatures of nonequilibrium spectral properties, and (3) to investigate electronic statistics and its non-thermal profile far from equilibrium.

\begin{figure}
\centering
\includegraphics[width=\linewidth]{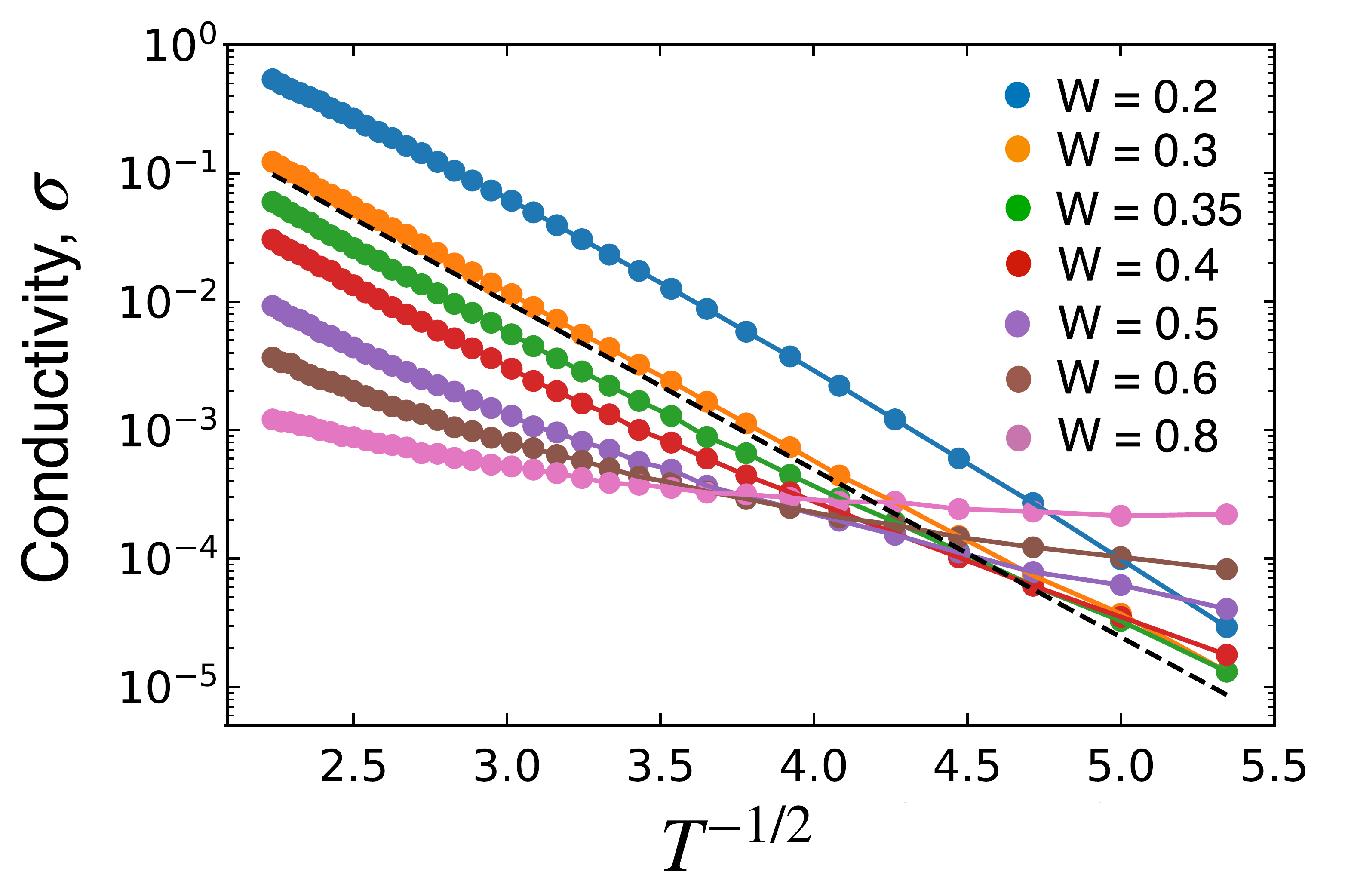}
\caption{Mott-$1/2$ Law. Numerically computed conductivity $\sigma$ plotted against $T^{-1/2}$ for various disorder strengths $W$ with the energy gap $\Delta = 0.3$ (all parameters described later in the text). We observe that the behavior approaches $\ln\sigma\sim -T^{-1/2}$ in the high-temperature limit and agrees with Mott's theory (straight dashed line) for the widest range of temperatures at $W=0.35$. The numerically computed conductivity deviates from Mott's relation in the large $W$ limit as the VRH argument is invalidated since too many levels get occupied. Detailed computational discussions on the lattice model will be given in later sections.}
\label{fig:1}
\end{figure}

To motivate the study, we first summarize the concept of variable range hopping (VRH) transport in equilibrium, following Mott's argument~\cite{mott1968conduction}. We consider electron transport through hops, ${\cal W}$, in disordered levels on a lattice. The probability of hops between nonlocal sites with the level difference $\Delta \epsilon$ depends on the spatial overlap between localized states separated by $R$ as, similar to the Miller-Abraham's expression~\cite{PhysRev.120.745},
\begin{equation}
    \mathcal{W} = \mathcal{W}_0 \exp \left[  - \frac{2R}{\xi} - \frac{\Delta \epsilon}{k_BT} \right]
    \label{mil-ab-eq}
\end{equation}
where $\xi$ is the localization length, $T$ the temperature with the Boltzmann constant $k_B$, and the overall constant ${\cal W}_0$. Mott proposed that the most probable hops are those that maximize the exponent in the hopping probability, effectively balancing the distance $R$ and the energy difference $\Delta \epsilon$. To achieve this, he proposed a statistical approach where the number of states within a $d-$dimensional sphere of radius $R$ and energy width $\Delta \epsilon$ is given as $Vg(\epsilon_\text{F})\Delta \epsilon$, where $V \sim R^d$ is the volume and $g(\epsilon_\text{F})$ is the density of states of disordered levels at the Fermi level $\epsilon_\text{F}$. Assuming that there is at least one state available to hop in this volume and the energy range, we relate the probable level spacing given by the range of hopping as
\begin{eqnarray}
    \Delta \epsilon \sim \frac{1}{g(\epsilon_\text{F})R^d}
\end{eqnarray}
Now substituting this term to Eq.~(\ref{mil-ab-eq}) and maximizing the exponent gives us a generalized equation for the conductivity which is also known as Mott's law of variable range hopping conduction
\begin{equation} \label{mott eq}
    \sigma = \sigma_0 \exp\left[- \left( \frac{T_0}{T} \right) ^\gamma\right]
\end{equation}
where $\gamma = 1/(d+1)$ and $T_0 = \alpha/(k_Bg(\epsilon_\text{F})\xi^d)$ with some numerical constant $\alpha$. For a one-dimensional system, the value of the exponent is $\gamma = 1/2$. This has been studied extensively using semi-classical resistor network approaches \cite{10.1063/10.0034343,lee1984variable,shklovskii2013electronic, EPSTEIN20019,Apsley1974-APSTFO} and shown experimentally in a wide variety of disordered systems \cite{doi:10.1021/nn101376u,PAASCH200297,evans2017mono}. In this work, we show that Mott's variable range hopping behavior emerges in our one-dimensional tight-binding chain, as the conductivity (obtained through Green's function calculations) typically develops the functional form of Eq~(\ref{mott eq}) in some disorder limit. We highlight our result for the conductivity, as a function of temperature in our model, in Fig.~\ref{fig:1} where the $\log(\sigma)$ varies linearly with $T^{-1/2}$. These results, along with our model and the relevant parameters, are described later in the text. 

In this paper, we present a simple quantum model consisting of an infinite tight-binding chain and on-site disorder potential to model VRH. We apply an electric field to this chain and compute the nonequilibrium Green's functions in the steady-state limit to study different spectral properties of the disordered system under bias. Since we consider a nonequilibrium steady-state, we introduce dissipators via an infinite fermionic reservoir coupled to each site to drain the energy injected by the electric field. We investigate the signatures of localization by studying the spectral function of a disordered lattice. By increasing the electric field, we demonstrate the localization-delocalization crossover. We also compare our lattice results with those of the coherent potential approximation (CPA) \cite{elliott1974theory,janivs2021dynamical, dohner2022nonequilibrium}, where we calculate the effect of disorder through an effective medium and incorporate it self-consistently into the lattice quantities. We demonstrate that the statistical spectra show unexpected behavior of the distribution edge receding in the opposite direction of the band gap. We finally discuss the localization of the wavefunction in an open nonequilibrium system.

The rest of the paper is organized as follows. In section \ref{sec:level2}, we describe our quantum mechanical model and discuss our calculation methods of Green's functions and transport properties. In section \ref{sec:level3}, we present our results. For reliable comparison, we briefly discuss the non-disordered lattice case in section \ref{sec:level3-1}. In section \ref{sec:level3-2} we discuss the disordered lattice case in equilibrium and out of equilibrium. Finally, in section \ref{sec:level4} we discuss our results and conclusions. 

\section{\label{sec:level2}Model and Method: Disordered Chain under DC bias}

\begin{figure}
\includegraphics[width=\linewidth]{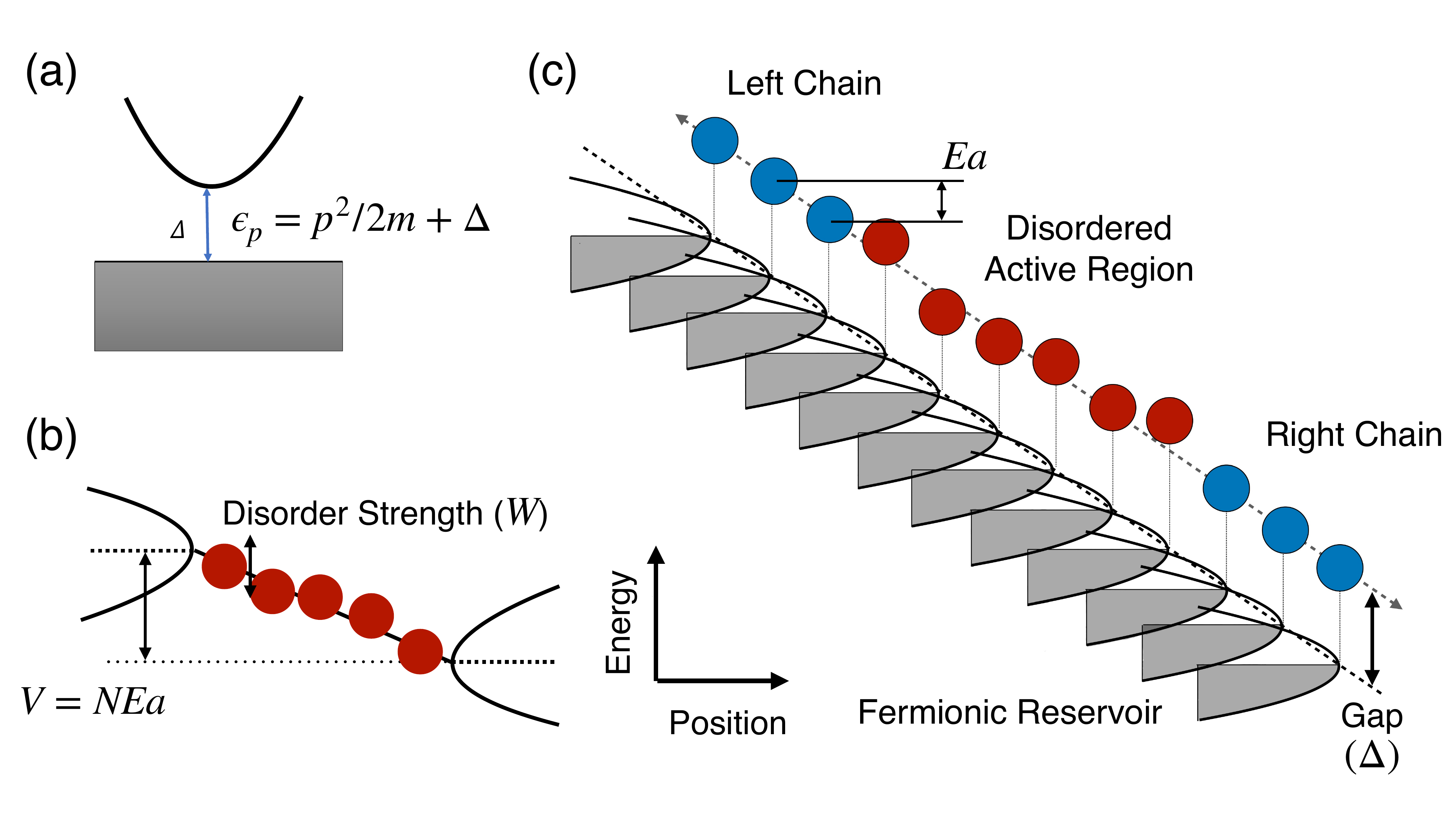}
\caption{(a) One-dimensional band separated by the gap $\Delta$ from the Fermi level in equilibrium. (b) Disordered chain in junction geometry. While commonly considered, this setup suffers spurious scattering from the band misfit at any electric field as we approach the bulk limit. (c) Infinite-chain setup with an electric field. Each site and its reservoir is subject to the electrostatic potential throughout the chain. The central region (red) is disorder-active with the random potential shift in the range $(-W,W)$.}
\label{fig:2}
\end{figure}

We aim to study the electron transport in the bulk limit. As shown in Fig.~\ref{fig:2}(a), the system is an insulator with dispersion relation $\epsilon_p=p^2/2m+\Delta$, with the insulating gap $\Delta$ removed from the Fermi energy (set to zero) in equilibrium. To implement the nonequilibrium system with an electric field $E$, one may consider a junction geometry as depicted in Fig.~\ref{fig:2}(b). However, this junction model leads to finite-size effects that are undesirable for bulk properties. With the number of sites $N$, and the lattice constant $a$, the bias $V=NEa$ eventually grows to the condition that the source and drain bands do not overlap, especially for a long-chain limit. This junction setup, although commonly used in mesoscopic physics, may interfere with bulk properties, and we do not discuss in this work.

We therefore construct the nonequilibrium lattice in the bulk limit so that the lattice is translationally invariant in the clean limit. As pictured schematically in Fig.~\ref{fig:2}(c), we start with an infinite one-dimensional tight-binding chain with a uniform electric field throughout the whole chain. We also introduce fermion reservoir chains that couple to each site of the main chain as shown. The role of the fermion reservoir is to drain the excess energy driven by the electric field and to establish a nonequilibrium steady-state~\cite{han2013solution}. The Hamiltonian reads as
\begin{eqnarray}
\mathcal{H} = && -t\sum_{l=-\infty}^\infty (d_l^{\dagger}d_{l+1} + d_{l+1}^{\dagger}d_{l})+ \sum_l \epsilon_l d_l^{\dagger}d_{l} \nonumber\\
&& 
+ \sum_{l\alpha} (\epsilon_{\alpha} - lE) c_{l\alpha}^{\dagger}c_{{l\alpha}} - \frac{g}{\sqrt{L}}\sum_{l\alpha} (c_{l\alpha}^{\dagger}d_{l} + h.c.).
\label{eq:1}
\end{eqnarray}
Each site of the main chain has the electron creation (annihilation) operator $d^{\dagger}_l$   $(d_l)$ mixing with the reservoir operators $c^\dagger_{l\alpha}$ ($c_{l\alpha}$) with the continuum index $\alpha$. $t$ is the tight-binding parameter and $\epsilon_l$ the site energy of the main chain. We set $t=1$ as the unit of energy. With the electric field $E$, the site energy at site $l$ is
\begin{equation}
    \epsilon_l = 2t + \Delta + V_l - l\cdot eEa,
\label{eq:2}
\end{equation} 
with the lattice constant $a=1$ set as the unit of length. The electric charge $e=1$ is set as the unit of charge. 

The Anderson disorder in the active region of the main chain is set by $V_l$ with 
\begin{equation}
    V_l=\left\{\begin{array}{ll}
    \mbox{random in }[-W,W] & \mbox{for }-N/2\leq l\leq N/2 \\
    0 & \mbox{otherwise}
    \end{array}
    \right..
\end{equation}
In this work, we choose $N = 501$ disorder active sites, which is large enough to study bulk properties and is computationally feasible to perform finite lattice calculations. The electric field is probed from the lowest value of $E = 10^{-4}$ up to $E = 0.1$ which typically corresponds to the range of $[10,10^4]\text{ kV/cm}$ which includes the typical experimentally relevant range.

The density of states for the reservoir chain, $L^{-1}\sum_\alpha\delta(\omega-\epsilon_{l\alpha})$ with the normalization $L$ for the length of the reservoir chain, is considered structureless and of infinite bandwidth, for simplicity. The electrons on the main chain hybridize with the reservoirs with the hybridization $\Gamma(\omega)=\pi g^2 L^{-1}\sum_\alpha\delta(\omega-\epsilon_{l\alpha})\equiv \Gamma$. The parameter $\Gamma$ accounts for the level broadening and also for the dissipation rate. The reservoirs are in thermal equilibrium at the bath temperature $T$, with the distribution at site $l$ given by the Fermi-Dirac function $f_{\rm FD}(\omega-lE)=[1+e^{(\omega-lE)/T}]^{-1}$ displaced by the electric potential. This mismatch of the reservoir levels due to the electric field results in broken detailed balance at local sites. By using the Keldysh Green's function method, the effect of the reservoirs can be exactly incorporated as discussed below~\cite{li2015electric}. 

\subsection{\label{sec:level2-1} Infinite Disordered-Lattice Calculation}

We numerically compute Green's functions (GFs) on an infinite chain with the disorder active on a finite range of length $N$. Inversion of an infinite matrix for the GF is avoided by introducing the disorder-inactive semi-infinite chains via self-energies. The GF over the disorder-active sites can be set up as the matrix inversion problem of an $N\times N$ matrix\cite{li2015electric} as
\begin{eqnarray}\label{eq:3}
    [\mathcal{G}^{R}(\omega)^{-1}]_{ij}  & = & (\omega - \epsilon_i + i\Gamma)\delta_{ij} + t\delta_{|i-j| = 1} \\
& & 
    - t^2 F^R_{-}(\omega - NE/2)\delta_{i=j=-N/2} \nonumber \\
& & 
    - t^2 F^R_{+}(\omega + NE/2)\delta_{i=j=N/2} \nonumber 
\end{eqnarray}
where the reservoir self-energy is denoted as $-i\Gamma$. The last two terms $t^2F^R_-$, $t^2F^R_+$ are the self-energies of the semi-infinite leads attached to the LHS and RHS of the active sites region (Fig. \ref{fig:2}), respectively. The retarded Green's function at the end site of the left/right semi-infinite chain, $F^R_{\pm}$, can be computed recursively \cite{li2015electric} as 
\begin{equation} \label{eq:Fret}
    F^{R}_{\pm}(\omega)^{-1} = \omega + i\Gamma - (2t + \Delta) - t^2F^{R}_{\pm}(\omega \pm E).
\end{equation}
The full Green's function in Eq.~(\ref{eq:3}) is computed numerically by inverting the matrix on the RHS with random diagonal terms. We similarly compute the lesser Green's function numerically using the steady-state Keldysh equations~\cite{haug2008quantum,han2013solution}
\begin{equation}
\mathcal{G}^{<}_{ij}(\omega) = \sum_{k=1}^N \mathcal{G}^{R}_{ik}(\omega) \Sigma^{<}_k(\omega) [\mathcal{G}^{R}_{jk}(\omega)]^{*} 
\label{eq:4}
\end{equation}
where the lesser self-energy at site $k$
is given as $\Sigma^{<}_{k}(\omega) = 2i\Gamma f_{\rm FD}(\omega+kE) + t^2\delta_{k,-N/2}F^{<}_{-}(\omega-NE/2) + t^2\delta_{k,N/2}F^{<}_{+}(\omega+NE/2)$ with $f_{\rm FD}(\omega) = [1+e^{\omega/T}]^{-1}$. The first term of the lesser self-energy describes the local contribution due to the Fermion reservoir and the last two terms describe the contribution from the semi-infinite chains. The lesser Green's function similarly as $F^R_{\pm}$ as 
\begin{equation} \label{eq:Fless}
    F^{<}_{\pm}(\omega) = |F^{R}_{\pm}(\omega)|^2[2i\Gamma f_{\rm FD}(\omega)
    + t^2F^{<}_{\pm}(\omega\pm E)].
\end{equation}
All GFs are computed at a fixed disorder configuration. The disorder-averaging is performed only after the GFs are computed, and no statistical assumptions are used in the calculation otherwise. We choose $N=501$ and the number of disorder configurations is over 4000. All results are well converged, and within this statistical spread, the solutions are numerically exact.

In the absence of disorder ($W=0$), the Green's functions satisfy the relation~\cite{li2015electric}
\begin{equation}
    {\cal G}^{R,<}_{i,j}(\omega)={\cal G}^{R,<}_{i+k,j+k}(\omega-kE),
\end{equation}
and the observable quantities are site-independent. With disorder, the configuration average restores this symmetry away from the edges of the disorder-active region.

\subsection{\label{sec:level2-2}Coherent Potential Approximation (CPA)}

To interpret the explicit lattice calculations, we complement them with a simplified approach from the coherent potential approximation (CPA). This method has been widely used to study disordered systems, over a wide range of problems. The CPA has been put into the context of the dynamical mean field theory (DMFT) \cite{aoki2014nonequilibrium, georges1996dynamical,10.3389/fphy.2020.00324} by using a local effective medium approximation. Here, the inhomogeneous spatial correlations induced by the random disordered potentials are replaced by a local effective medium potential, described as the CPA self-energy, $\Sigma_{\text{CPA}}(\omega)$. This CPA self-energy is determined self-consistently by embedding an impurity in the effective potential and by computing the average disordered Green's function for multiple impurity configurations and mapping on the effective medium Green's function. 
The numerical CPA self-consistent method is summarized as  
\begin{enumerate}
\item{We start with an initial guess for the CPA self-energy as $\Sigma_{\text{CPA}}(\omega)$ often chosen as $\Sigma_{\text{CPA}}(\omega)=0$, we compute the Green's function of the semi-infinite leads recursively~\cite{li2015electric} with an electric field $E$. The retarded Green's function is defined as follows.
\begin{eqnarray} \label{eq:cpares}
[\bar{F}^{R}_{\pm}(\omega)]^{-1} & = & \omega + i\Gamma-2t-\Delta \\  & & - \Sigma^{R}_{\text{CPA}}(\omega) - t^2\bar{F}^{R}_{\pm}(\omega \pm E) \nonumber
\end{eqnarray}
and the lesser Green's function is defined as
\begin{eqnarray}
    \bar{F}^{<}_{\pm}(\omega) & = & |\bar{F}^{R}_{\pm}(\omega)|^2 [2i\Gamma f_{\rm FD}(\omega) \\ & &+ \Sigma^{<}_{\text{CPA}}(\omega) + t^2\bar{F}^{<}_{\pm}(\omega \pm E)]. \nonumber
\end{eqnarray}
Note that the GF $\bar{F}^{R,<}_\pm(\omega)$ contains the CPA self-energy $\Sigma_{\rm CPA}(\omega)$ and  differs from the disorder-free GF $F^{R,<}_\pm(\omega)$ in the previous subsection, Eq.~(\ref{eq:Fret}).
We define the total lead-Green's function as : $\bar{F}^{R,<}(\omega) = \bar{F}^{R,<}_{+}(\omega+E) + \bar{F}^{R,<}_{-}(\omega-E)$ which represents the self energy of the effective medium. The nonequilibrium effects of the electric field are incorporated in a similar fashion as the lattice calculations in the semi-infinite chain up to the single impurity sites by shifts in the frequency \cite{li2015electric}.}

\item{We compute the retarded Green's function excluding the disorder at the central site : 
\begin{equation} \label{eq:eff-Ret}
\mathcal{G}_{0}^{R}(\omega) = [\omega + i\Gamma-2t-\Delta - t^2\bar{F}^{R}(\omega)]^{-1}
\end{equation}
similarly, the lesser Green's function is calculated as:
\begin{equation} \label{eq:eff-Less}
    \mathcal{G}_{0}^{<}(\omega) = |\mathcal{G}_{0}^{R}(\omega)|^2[2i\Gamma f_{\rm FD}(\omega) + t^2\bar{F}^{<}(\omega)]
\end{equation}
The medium Green's function defined in Eqs. (\ref{eq:eff-Ret}) and (\ref{eq:eff-Less}) is mathematically analogous to the Weiss GF, commonly encountered in DMFT. Physically, this represents a single-site problem embedded in a self-consistent effective environment which is incorporated using the self-energy of the leads $\bar{F}^{R,<}(\omega)$. }

\item{We compute the disorder-averaged retarded Green's functions : 
\begin{equation} \label{eq:gfcpa}
G^{R}_{\text{loc}}(\omega)  = \langle [{\cal G}^R_0(\omega)^{-1} - V]^{-1} \rangle_{V}
\end{equation}
and lesser Green's functions
\begin{equation}
    G^{<}_{\text{loc}}(\omega) = |G^{R}_{\text{loc}}(\omega)|^2 [2i\Gamma f_{\rm FD}(\omega)
     + t^2\bar{F}^{<}(\omega)] \nonumber
\end{equation}
Disorder averaging in the CPA calculation is defined as $\langle f \rangle_V = \int_{-W}^{W} P(V) f(V)dV$ where $P(V) = (2W)^{-1} \Theta(W-|V|)$ is the uniform distribution function of disorder potentials. A detailed discussion of an analytic procedure at zero field is given in Appendix~\ref{App:2}. 
}
\item{We compute the CPA self-energy
\begin{equation} \label{eq:cpaSelf}
\Sigma^{R}_{\text{CPA}}(\omega) = \mathcal{G}^{R}_{0}(\omega)^{-1} - G^R_{\text{loc}}(\omega) ^{-1}
\end{equation}
\begin{equation*} 
\Sigma^{<}_{\text{CPA}}(\omega) = \frac{ G^{<}_{\text{loc}}(\omega) }{| G^{R}_{\text{loc}}(\omega) |^2} - \frac{ \mathcal{G}_{0}^{<}(\omega)}{|\mathcal{G}_{0}^{R}(\omega)|^2}
\end{equation*}
}
\item{We repeat the above process starting from step 1 until we get a converged solution.}

\end{enumerate} 

In the next section, we present our results starting with the equilibrium disorder-free case, and we briefly summarize the effect of dissipation $\Gamma$ and temperature $T$. Following that, we present the spectral function of the disordered lattice and compare our results from the lattice calculation and CPA method. In the nonequilibrium case, we discuss the behavior of the Lifshitz tail with the electric field and its effect on localization in the 1D disordered-lattice by numerically studying the inverse-participation ratio (IPR). 

\section{\label{sec:level3}Results and Discussions}

\subsection{\label{sec:level3-1}Disorder-Free Case ($W = 0$)}
\subsubsection{Effect of Damping in Gapped Insulators}

We discuss our system in the zero disorder limit and highlight the effect of dissipation $\Gamma$ on the conductivity of the gapped materials. We rigorously compute the effect of damping on the conductivity and the occupation number in the low-temperature limit. An analytic expression for occupation number is computed from the lesser Green's function as $n_{p}(t) = -iG^{<}_p(t,t)$, as detailed in Appendix \ref{App:1}. In the limit of $E \rightarrow 0$ the charge excitation per site can be calculated as
\begin{equation}
n_{\rm loc} = \frac{\Gamma}{2\pi}\sqrt{\frac{2m}{\Delta}}
\label{n_loc}
\end{equation}
where $m$ is the effective mass of the system and $\Delta$ denotes the gap parameter as mentioned in the previous section. The effective mass $m$ in the continuum model is related to the tight-binding parameter $t$ by $ta^2=\hbar^2/(2m)$.
We can similarly compute the current as  
\begin{equation*}
    J = \int \frac{d\bar{p}}{2\pi} \frac{\bar{p}}{m} n_{\bar{p}} = \frac{\Gamma^2E}{16\pi m} \frac{\sqrt{2m\Delta}}{\Delta^3}
\end{equation*}
which gives us the DC conductivity as
\begin{equation} \label{eq:10}
\sigma = \frac{\Gamma^2}{16\pi}\frac{\sqrt{2m\Delta}}{m\Delta^3}
\end{equation}
Fig.~\ref{fig:6} (in Appendix~\ref{App:1}) shows the variation of the occupation number and conductivity with damping $\Gamma$ and presents a comparison between our numerical calculation and the analytic derivation. We calculate the mobility ($\mu=\sigma/n_{\rm loc}$) and observe that it increases linearly with the damping $\Gamma$. This is in contrast with the low-field Drude limit result, $\sigma\propto \tau \propto 1/\Gamma$, which was shown in \cite{han2013solution} for half-filling. In our model, however, the main chain is kept above the Fermi level by $\Delta$ and is scarcely occupied. The lattice is occupied only for a fraction of time proportional to $\Gamma/\Delta$ as shown in Eq.~(\ref{n_loc}), and the time of acceleration due to the electric field is also during the time of occupation, leading to the drift velocity $\propto E\Gamma$. Hence, the conductivity increases as $\Gamma^2$ in the absence of disorder. 

\subsubsection{Equilibrium Case: High-temperature limit}
We next study the temperature dependence of transport quantities like the occupation number $n_{\rm loc}$ and the conductivity $\sigma$. The occupation number in the equilibrium limit is obtained from the retarded Green's function. The occupation number at the wave number $k$ in equilibrium is given as
\begin{equation} \label{eq:5}
n(k,\omega) = 2i\Gamma f_{\rm FD}(\omega) |\mathcal{G}^{R}_k (\omega)|^2
\end{equation}
which is integrated over all $\omega$ and $k$ values to obtain the complete expression for occupation as
\begin{equation} \label{eq:12}
n(T) \approx  \sqrt{2mT} e^{-\Delta/T}.
\end{equation}
It is also verified numerically in the high-temperature limit that the occupation of the main chain is independent of the dissipation. Electrons are thermally excited to high enough energies to cross the gap, and hence the occupation increases. 

In equilibrium, the conductivity has an activation-like behavior (Arrhenius behavior~\cite{Yuen2009}). The conductivity has the empirical form
\begin{equation}
\sigma(\Gamma,T) = \sigma_0(\Gamma) e^{-\Delta/T}
\end{equation}
The pre-factor depends on $\Gamma$ as $\sigma_0 \sim \Gamma^{-1}$ which recovers the Drude behavior of conductivity in the high-temperature limit. Thermal excitation is the main source of scattering, and hence only higher energy electrons contribute to the scattering process. As the temperature is lowered, the conductivity deviates from the typical activation behavior since there is lower thermal excitation and electrons do not have enough energy to occupy the band. Hence, the few energetic electrons are more likely to be scattered into the reservoir and our behavior coincides with Eq.~(\ref{eq:10}).

\begin{figure*}
\centering
\includegraphics[width=\linewidth]{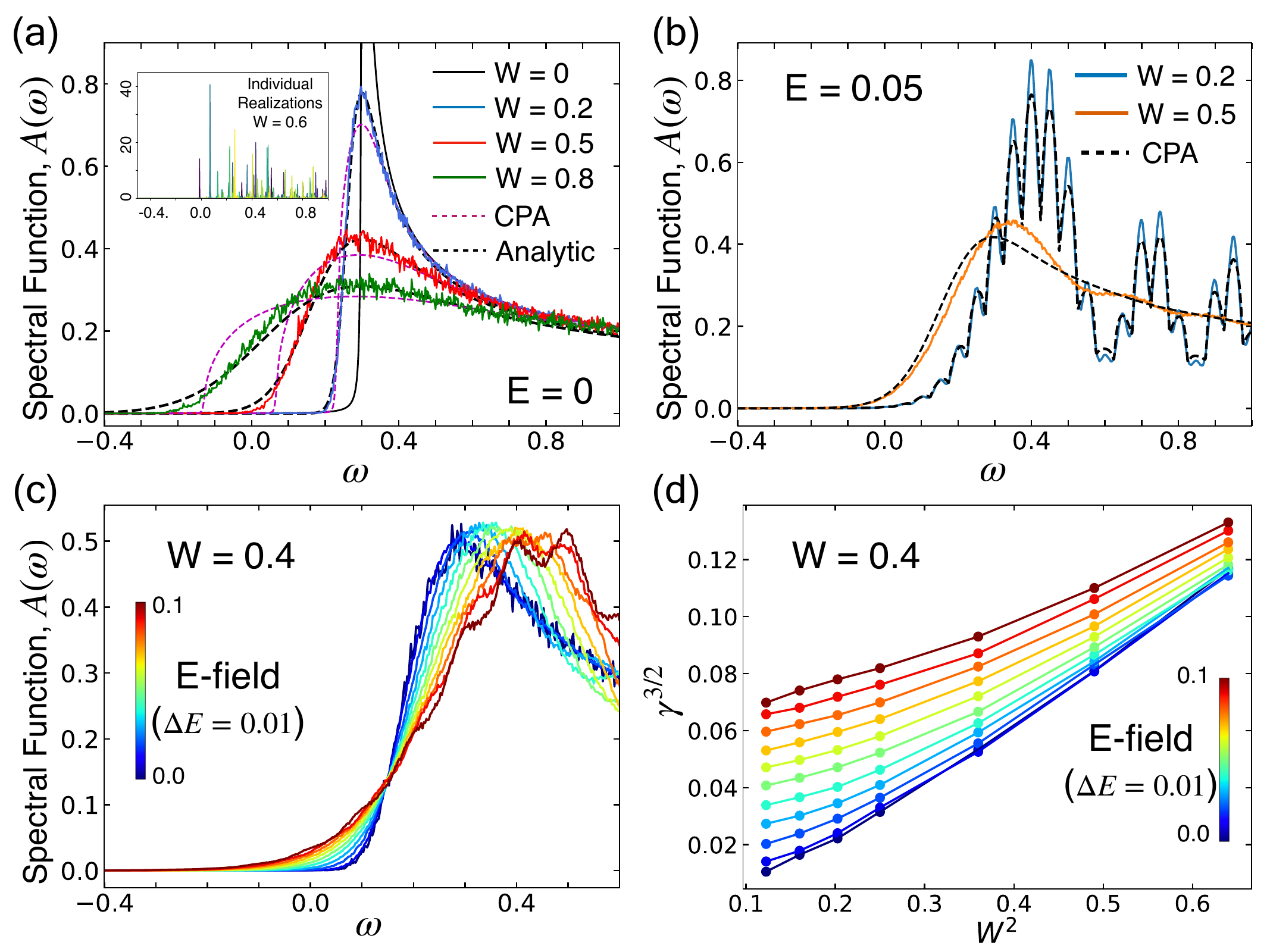}
\caption{(a) Disorder-averaged spectral function for different values of the disorder $W$ at zero electric field $E = 0$ with gap $\Delta = 0.3$, $\Gamma = 0.0005$ and $T = 0.01$ . Pink dashed lines correspond to the spectral function obtained from the CPA calculation, which shows a fair match with the lattice calculation but shows no exponential Lifshitz tails. Individual disorder configurations are plotted for $W =0.6$ in the inset, which shows distinct localized states. The black solid line is the $1-d$ tight-binding density of states inside the band. (b) Disordered average spectral function at $E = 0.05$ for weak ($W=0.2$) and strong ($W=0.6$) disorder. Under the electric field, the system develops slow oscillations, which is a modified Airy function, and the fast oscillations are the Wannier Stark peaks occurring at the frequency spacing of $\Delta\omega = E$.  At strong disorder, there are no such peaks, indicating that the electric field is not strong enough to break localization. (c) Progression of the Lifshitz tail with electric field at disorder $W = 0.4$ for varying electric fields. Starting from the dark-blue curve at zero electric field to the red curve at electric field $E=0.1$ with the spacing $\Delta E = 0.01$, the electric field smears the band edge which superposes onto the smearing due to disorder. (d) The decay width of the Lifshitz tail $\gamma$ as a function of $W^2$.}
\label{fig:12}
\end{figure*} 

\subsection{\label{sec:level3-3}Mott's $1/2$-Law at Zero E-field}
We discuss the linear transport at very low bias (numerically electric field is kept at $E = 5\times 10^{-5}$), and verify Mott's law as previously shown in Fig. \ref{fig:1}. In our lattice calculation, we calculate the local current $J$ at each site \cite{li2015electric} using the lesser Green's function computed from Eq. (\ref{eq:4})
\begin{equation}
J_l = -t \int [\mathcal{G}^{<}_{l,l+1}(\omega) - \mathcal{G}^{<}_{l+1,l}(\omega)]\frac{d\omega}{2\pi}
\label{eq:jloc}
\end{equation}
The conductivity is computed as $\sigma = \langle J_l\rangle/E$, which is averaged over many disorder configurations and also over multiple sites to obtain an averaged local conductivity of the lattice. In the absence of disorder, $J_l$ is identical on all sites. Fig.~\ref{fig:1} demonstrates that $\log(\sigma)$ decreases with $T^{-1/2}$, consistent with the variable range hopping behavior ($\log(\sigma)\propto -T^{-1/2}$ shown as the dashed line for comparison). This provides an important benchmark where Mott's behavior emerges from a purely quantum mechanical calculation of the Green's function theory without any reference to Mott's statistical argument. In the low-temperature limit, the behavior deviates significantly because of the hybridization $\Gamma$ with the fermionic bath. 

The Mott behavior as seen in Fig.~\ref{fig:1} also varies with the disorder strength as it is observed that the plot deviates from Mott's scaling (dashed line) both in the high- and the low-disorder limit. At $W \sim \Delta$, $\log(\sigma)$ has the most linear behavior with respect to $T^{-1/2}$. At low disorder strength, the conductivity has slight downward concavity which highlights that conductivity assumes the Arrhenius behavior ($1/T$-dependence).  However, at higher disorder strength the conductivity becomes independent of temperature as the levels fall deep in the Fermi sea and localization is too strong for the electrons to hop around. This allows us to identify the VRH regime in our model which typically corresponds to $W \sim \Delta$ in our quantum mechanical calculation of our disordered-lattice model. In this limit, the localized levels appear close to the Fermi level which may allow electrons to hop into the unoccupied band. The range of the VRH regime will be further discussed in the following sections.

The fermion reservoirs, incorporated exactly in the Green's function theory, provide energy relaxation when they are placed at different chemical potentials through the particle exchange of hot and cold electrons between the system and the reservoirs. The spectrum for the exchange, the particle-hole continuum, has the same dynamics as an Ohmic bath of acoustic phonons, and mimics the energy reservoirs used in Mott's VRH. It is important to note that this purely electronic model, exactly solvable numerically, is shown to be consistent with the vast literature on the topic and thus provides a platform to study nonlinear electronic transport.

\subsection{\label{sec:level3-2}Spectra of Disordered Lattice under E-field}

One of the goals of this study is the discussion of localization in a disordered tight-binding chain and the effect of an electric field on Anderson localization. It is known that in a 1D system, any amount of disorder can lead the system into localization \cite{thouless1974electrons, Mott01041961}. The averaged spectral function is computed from the imaginary part of the retarded GF
\begin{equation}
A(\omega)=-\frac{1}{\pi}{\rm Im}\,\langle G^R_{ii}(\omega)\rangle,    
\end{equation}
over the disorder average and plotted in Fig.~\ref{fig:12}. Averaging over multiple disorder configurations, the spectral function becomes smooth and the band edge shifts. We observe not only the broadening of the band with increasing disorder but also the exponential Lifshitz tail appearing along the edge of the band. Our numerically computed spectral function is an accurate match for the analytic expression of the spectral function [depicted as the black curves in Fig. \ref{fig:12}(a)] reported in previous studies \cite{RevModPhys.64.755}. The inset shows individual disorder realizations for some disorder strength which shows localized levels marked by $\delta$-function-like spikes in the DOS.

We compare our disordered-lattice calculations with the CPA calculation (shown in pink dashed lines) at zero field in Fig.~\ref{fig:12}(a). The disorder-averaged spectral function shows remarkable similarity to the CPA calculation. This is an important benchmark for both the finite lattice and the CPA calculations as we can highlight some key aspects of both approaches. CPA as a single-site approximation replaces the inhomogeneity of a disordered lattice with an averaged coherent potential. Hence, it does not show any Anderson localization but highlights the broadening of the band as a result of the scattering of the complex potential. In contrast to the CPA-averaged spectral function, the disordered-lattice spectral function forms a Lifshitz tail rather than a sharp band edge. In the high-field limit, however, the non-local effects of the disorder become disentangled, the CPA becomes more reliable, and its agreement with the disordered-lattice calculations is excellent as shown in Fig.~\ref{fig:12}(b).

\begin{figure}
\includegraphics[width=0.48\textwidth]{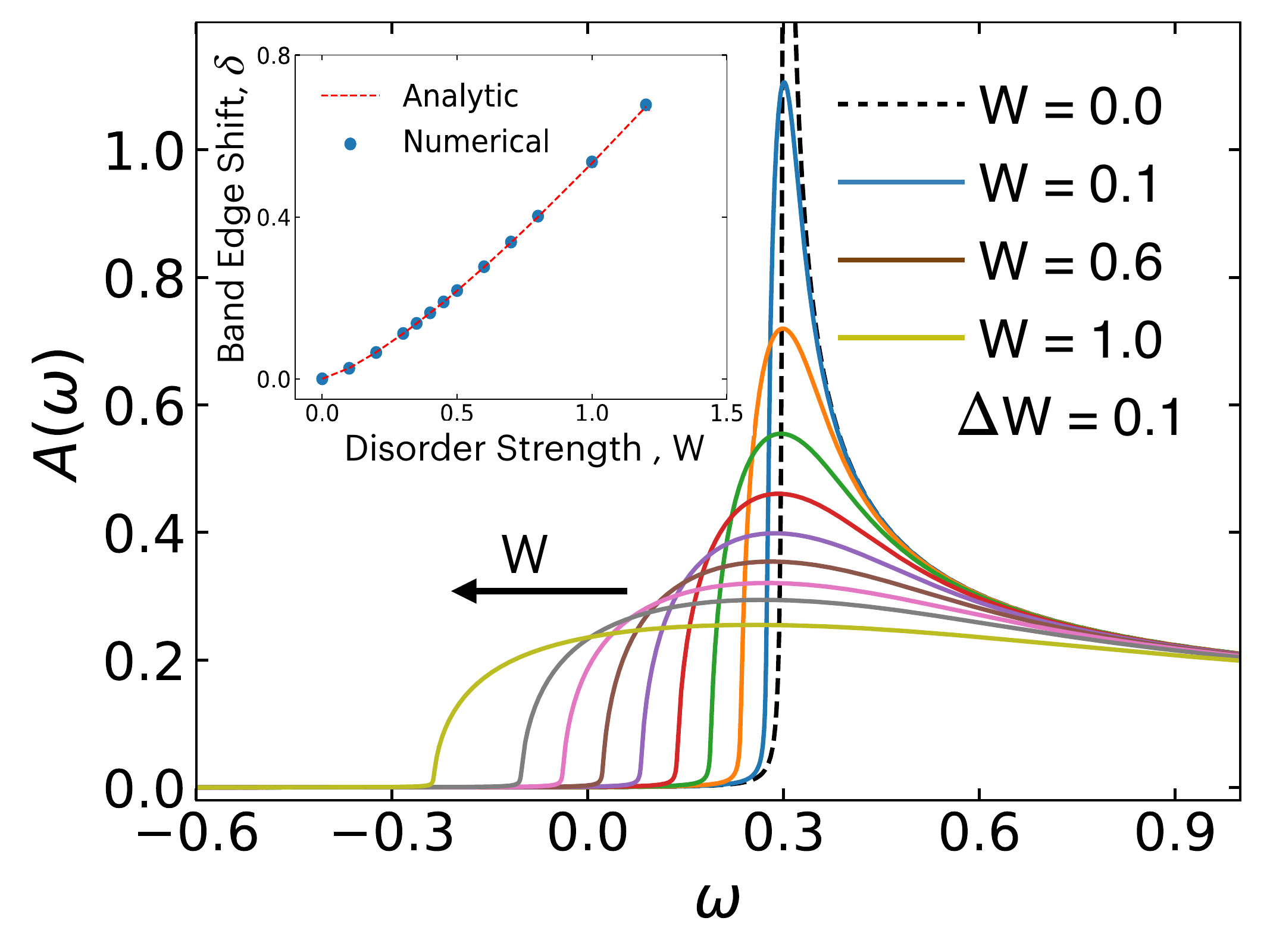}
\caption{CPA spectral function at $\Gamma = 0$ and $E = 0$ with varying disorder strengths. In the inset, the band edge shift is compared between the disordered-lattice calculation and the analytical CPA results.} 
\label{fig:10}
\end{figure} 

In Fig.~\ref{fig:12}(b-d), we show the effect of the electric field on the spectral properties of the disordered chain. Applying the electric field to the disordered chain results in the spectral function developing oscillations, which indicate the tendency for the system to get delocalized. At a very strong electric field and low disorder strength, Anderson localization crosses over to weak localization, which is depicted as Wannier Stark peaks in Fig.~\ref{fig:12}(b) occurring with the frequency interval of $\Delta\omega = E$. This is in agreement with some of the earlier works \cite{kirkpatrick1985localization} which have mentioned that the electric field delocalizes the system.

Fig.~\ref{fig:12}(b) shows two oscillation periods in the spectra. The slow oscillation represents the Airy function envelope, as can be seen in a continuum model. The Airy function envelope decays near the band edge as the electronic wave functions smear into the energetically forbidden zone. This spectral behavior, coincidentally, is of similar mathematical form to the exponential decay of the Lifshitz tail due to the disordered potential~\cite{halperin1965green,garcia2024fluctuation}. Applying the electric field further extends the Lifshitz tails into the gap as seen from Fig.~\ref{fig:12}(c). Combining the two limits of $(W>0, E=0)$ and $(W=0, E>0)$, we propose an empirical expression for this exponential Lifshitz tail for non-zero disorder and electric field as
\begin{equation}
    A(\omega) \sim \exp\left[ - \left| \frac{\omega-\Delta+ \delta}{\gamma} \right|^{3/2} \right]
    \label{spec:tail}
\end{equation}
below the band edge, where $\gamma$ is the decay width and $\delta$ represents the band edge shift. We discuss the behavior of both of these variables for varying disorder strengths and highlight the key underlying physics. 

\begin{figure*}
\includegraphics[width=\linewidth]{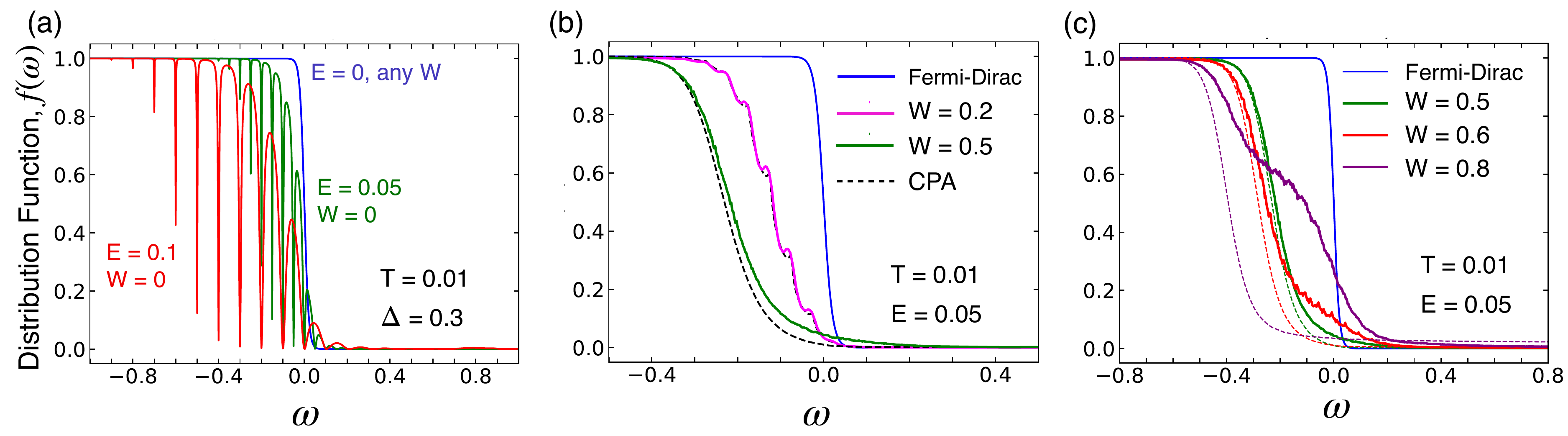}
\caption{(a) Distribution function of a clean system under an electric field at $T=0.01$, $\Gamma = 0.0005$ and $\Delta=0.3$. At zero field (blue), the fluctuation-dissipation limit is obeyed for all disorder strengths $W$. With a DC field, the distribution edge shifts to negative energy, on the opposite side of the band edge $\Delta>0$. This is due to the downstream depletion of electrons due to the energy flux from the field.
(b) Nonequilibrium distribution function under an electric field ($E =0.05$) for weak disorder. The down-shift of the distribution is enhanced by disorder due to the weakened localization by the field in the forward direction. The coherent-potential approximation shows excellent reproduction of the smeared Wannier-Stark ladder and the distribution shift.
(c) For stronger disorder, the distribution edge tends to shift back to the original Fermi level at $\omega=0$, possibly due to strong disorder overcoming the effect of the field.} 
\label{fig:11}
\end{figure*}

From the earlier analytic studies of Lifshitz tails \cite{halperin1965green, garcia2024fluctuation}, the decay of the Lifshitz tail $\gamma$ scales with respect to the $W$ as $\gamma^{3/2}\propto W^2$ at $E=0$ (blue-black) as shown in Fig. \ref{fig:12}(d). At higher electric fields, $\gamma$ tends to deviate from this scaling behavior. At zero disorder and at a finite field, the spectral tail is given in terms of the Airy function solution, $\gamma\propto E^{2/3}$ \cite{schwinger2001quantum}. Therefore, the tail length $\gamma$ is proposed to the form
\begin{eqnarray}
    \gamma = \max \left[\left( \frac{3W^2}{16\sqrt{t}} \right)^{2/3} , \left(\frac{3eEa\sqrt{t}}{4}\right)^{2/3}  \right].
\end{eqnarray}
The extension of the Lifshitz tail implies that randomly occurring localized levels would penetrate into the Fermi sea, facilitating electron excitations into the main chain, and would affect the transport properties of the disordered systems.  

The bandshift $\delta$ is important to understand the criterion for the VRH regime, and it is well captured by the CPA method. This is depicted in Fig.~\ref{fig:10} where the band edge shifts to lower energy with increasing disorder strength. We analytically derive the band edge shift at zero field using the local averaged Green's function, $G^R_{\rm loc}(\omega)$ in Eq.~(\ref{eq:gfcpa}) as a function of the disorder strength, which is shown in the Appendix~\ref{App:2}. In Fig.~\ref{fig:10}, we show that the band edge crosses the Fermi level at disorder $W \approx 2\Delta=0.6$, where this system typically develops metallic characteristics. In the disordered-lattice calculations, the crossing happens at a slightly lower value $W^{*} \sim 0.5$ due to the Lifshitz tail crossing over the Fermi level. This discrepancy for the threshold disorder becomes more important in transport behavior.

\subsection{\label{sec:level3-3}Distribution of Disordered Lattice under E-field}

While the retarded Green's functions give us the information on the single-electron dynamical response, such as the spectral functions, the lesser Green's functions provide us with the statistical information of the nonequilibrium steady-state limit. In particular, as will be demonstrated below, the electronic band structure contributes in a nontrivial manner that cannot be predicted by classical statistical network theories. The local distribution function at site $i$ is defined as 
\begin{equation}
   f_i(\omega) = \frac{-\text{Im}[{\mathcal{G}_{ii}^<}(\omega)]}{2\text{Im}[\mathcal{G}_{ii}^R(\omega)]},
   \label{eq:fi}
\end{equation}
with ${\cal G}_{ii}^<(\omega)$ computed from Eq.~(\ref{eq:4}). The distribution function $f_{i=0}(\omega)$ reduces to the Fermi-Dirac distribution function $f_{\rm FD}(\omega)$ in the zero-field limit ($E=0$) regardless of the disorder $W$ and obeys the fluctuation-dissipation theorem, as shown in Fig.~\ref{fig:11}(a). As the field $E$ is turned on in the clean limit ($W=0$), the distribution function (green and red curves) shows an unexpected trend in which the distribution edge shifts to negative frequency in the opposite direction of the band edge at $\Delta=0.3$. To understand this striking behavior, we rewrite the distribution function, Eq.~(\ref{eq:fi}), as
\begin{equation}
    f_i(\omega) = \frac{\sum_{k=-\infty}^\infty|{\cal G}^R_{ik}(\omega)|^2 f_{\rm FD}(\omega+kE)}{\sum_{k=-\infty}^\infty|{\cal G}^R_{ik}(\omega)|^2},
    \label{eq:fi2}
\end{equation}
which states that the steady-state nonequilibrium distribution is an average of the displaced Fermi-Dirac functions weighted by the response $|{\cal G}^R_{ik}(\omega)|^2$. The factor $|{\cal G}^R_{ik}(\omega)|^2$ weighs how much the source distribution at $k$ travels to the observation site $i=0$. 

With an electric field applied to a band above the Fermi energy, tunneling to upstream sites ($k<0)$ is reduced due to the potential barrier, while transport to the downstream sites is aided by the field. Therefore, the holes from $k>0$ contribute to the depletion of the distribution and to the downward shift of the distribution function. The spikes are due to the Wannier-Stark effect, also present in the spectral function.

Introducing disorder into the mix enhances the tendency of the negative shift of the distribution edge, as shown in Fig.~\ref{fig:11}(b). The Fermi-Dirac function in the zero electric field limit (blue) is shown for reference. For disorder strengths $W=0.2$ and $0.5$, the negative shift increases with $W$ further from the clean limit. Here, the disorder strength is kept moderate so that the Lifshitz tail, as shown in Fig.~\ref{fig:11}(a), has not crossed the Fermi energy. The strong field pumps energy into electrons moving in the direction of the field, and helps overcome the localization by disorder. Therefore, the imbalance between the upstream and downstream becomes amplified, and the downshift of the edge is enhanced. This confirms what we observe from the spectral function at high fields, which indicates that a strong electric field delocalizes the system, also corroborating earlier works \cite{prigodin1980one, kirkpatrick1986anderson}. The CPA results (black dashed lines) are in excellent agreement with the full lattice calculations, reproducing the smeared Wannier-Stark ladder and the shift of the distribution function edge.

For stronger disorder strengths where the Lifshitz tail begins to dip into the Fermi energy, the distribution function undergoes another unexpected change, as observed in Fig.~\ref{fig:11}(c). The distribution tail is strongly enhanced, reversing its previous trend, and partly recovers the distribution near the original Fermi energy, with the overall shape strongly departing from the Fermi-Dirac function. Although we only have an incomplete understanding of the mechanism, we may speculate as follows. (1) The enhancement of electron occupation indicates more electron flux from the upstream which has been assisted by the VRH mechanism. Interestingly, the CPA in this regime fails to capture this behavior. This leads us to suspect that (2) the correlation between impurities could be crucial to enhance the wavefunction overlap $|G^R_{ik}(\omega)|^2$ in Eq.~(\ref{eq:fi2}) near the Fermi energy.

We emphasize that the nonequilibrium electronic structure calculations, through the Keldysh Green's function technique, are essential to gain a realistic understanding of electronic transport. The rich physics seen above would not have been predicted without explicit consideration of the electronic lattice, which is often ignored in statistical network models. In the steady-state nonequilibrium, all electrons that have traveled from the remote past encode the band structure information and are counted for a non-thermal nonequilibrium distribution through the lesser Green's function. The shift of the distribution edge does not occur in particle-hole symmetric models~\cite{han2013solution,hanPRB2018}.

\subsection{\label{sec:level3-4}Wavefunction Localization under E-field}

In this subsection, we address the wavefunction localization by an electric field. To this end, we compute the inverse participation ratio (IPR) \cite{PhysRevLett.117.146601}, which is typically defined as the fourth power of a normalized eigenfunction $\phi_i$ summed over all spatial indices $i$ as
\begin{equation*}
    \text{IPR} = \sum_i^N |\phi_i|^4 
\end{equation*}
for a system of size $N$. This method gives some insights into the spatial localization or distribution of wavefunctions in a lattice. The IPR for an extended state in 1D is typically $\sim 1/N$, and it increases as the localization increases. In an open system, we redefine the IPR through the Green's function as 
\begin{equation}
    \text{IPR} = \frac{\sum_{x=1}^M |\mathcal{G}^R(x,\omega = -xE)|^4}{(\sum_{x=1}^M |\mathcal{G}^R(x,\omega = -xE)|^2)^2}
\end{equation}
where $\mathcal{G}^R(x,\omega)$ is the local retarded Green's function at position $x$ as computed in the Eq.~(\ref{eq:3}). The denominator provides the normalization and makes the IPR dimensionless. The IPR values are computed at the local Fermi level, which provides the main contribution to the transport.

\begin{figure}
\centering
\includegraphics[width=0.49\textwidth]{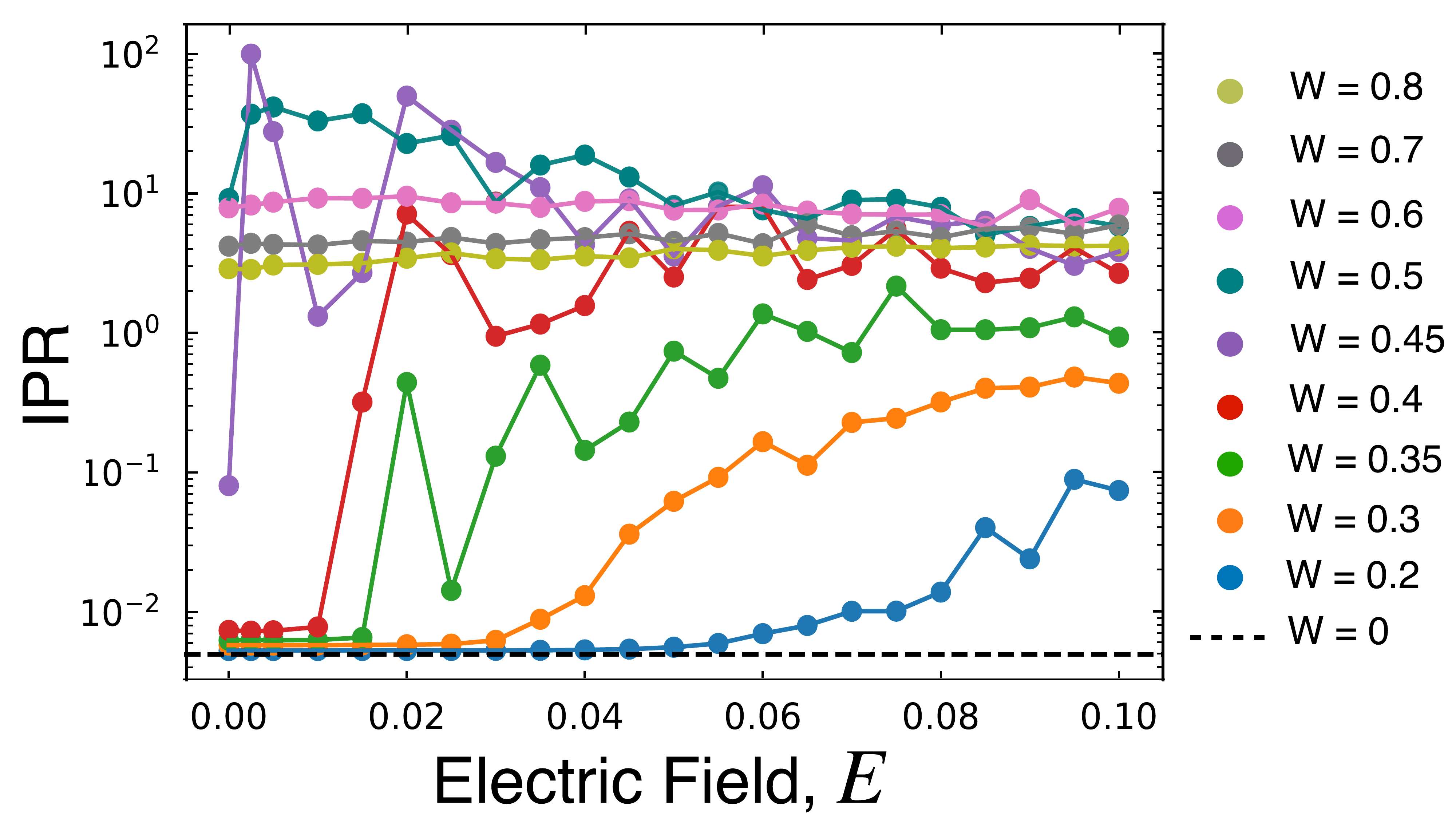}
\caption{Inverse-participation-ratio (IPR) plotted against electric field $E$. $\Gamma=0.0005$, $N=501$, $t=1$, $T=0.01$. The dashed line represents the $\text{IPR} \sim 1/N $ at zero disorder. Observe that in the disorder limit $\Delta < W < W^*$, the IPR at the Fermi level shows a sharp rise with $W$ as the electric field allows localized levels to penetrate into the Fermi sea. After the sharp rise, the electric field gradually delocalizes the system. At $W > W^*$, the IPR is independent of the electric field, as localization is far too strong to be affected by the electric field.}
\label{fig:14}
\end{figure}

Fig. \ref{fig:14} shows the evolution of the IPR as a function of the electric field at different disorder values $W$. It is observed that in the zero disorder limit (shown as a dashed line), we recover the $1/N$ behavior, showing that the states are all extended. At low disorder, the IPR increases gradually as the electric field increases. We observe weak localization of electronic levels, and the exponential part of the density of states smears into the gap and falls into the Fermi level due to the effect of the electric field. Now, increasing the disorder strength in the range
\begin{equation}
    \Delta < W < W^*,
    \label{wrange}
\end{equation}
the gradual rise in IPR becomes a much sharper increase with $W$ as the strongly localized levels are detected at the Fermi level. The upper limit $W^*=0.5$ (at $\Delta=0.3$) coincides with the condition that the Lifshitz tail crosses the Fermi energy, as discussed in Fig.~\ref{fig:12}. The CPA gives a slight overestimation of $W^*$ at $W^*\approx 2\Delta$. Near $W\approx W^*$, the IPR gradually decreases with increasing electric field, which indicates gradual delocalization due to the electric field. This is in agreement with what has been mentioned in earlier works \cite{prigodin1980one, kirkpatrick1986anderson} for the case of a discontinuous disorder. In the strong disorder limit, $W > W^*$, localization is too strong to be broken by the electric field, as observed by the flattening of the IPR with respect to the electric field.

\section{\label{sec:level4}Conclusion}
In this work, we analyzed the spectral properties of a disordered tight-binding chain under a DC electric field and coupled it to a dissipative bath, where the Fermi level is kept below the band edge. When disorder is present in the system, we observe typical Anderson localization. The spectral behavior of the localized system is characterized by solving an infinite one-dimensional lattice with a finite disorder-active region. 

We first verify Mott's law of the variable-range-hopping conductivity in the linear transport regime for our quantum mechanical model. This is one of the key results of this work, as we can obtain Mott's phenomenological model from quantum mechanical principles without resorting to statistical arguments. The lattice calculation, in the strong electric field regime, successfully captures exponential Lifshitz tails, which indicate the presence of localized levels beyond the allowed potential fluctuation bounds. We find that the electric field delocalizes a disordered system and allows for the system to crossover from Anderson localization to the Bloch oscillations. The advantage of this work is that the model is exactly solvable numerically, and with the verification of Mott's scaling and the Lifshitz tail, the model serves as a platform for future nonequilibrium studies.

Calculation of the local distribution function has been quite revealing. The unexpected shift and the non-thermal evolution of the nonequilibrium distribution from a minimally constructed lattice model demonstrates that a proper understanding of disordered transport in electronic lattices must include quantum mechanical bandstructure effects, going well beyond the statistical model. While more work is necessary to fully understand the mechanism for the rich behavior, it points to the competition of localization by disorder and delocalization by the field.

\begin{acknowledgments}
We thank J. P. Bird, J. Hofmann, G. Sambandamurthy, M. Randle, H. Zeng, and Z. Zhang for helpful discussions. Enlightening discussion with B. I. Shklovskii is especially acknowledged. We acknowledge computational support from the CCR at University at Buffalo. KM and HFF acknowledge the support from the Department of Energy, Office of Science, Basic Energy Sciences, under grant number DE-SC0024139. 
\end{acknowledgments}

\appendix

\section{\label{App:1} Linear Transport in a Clean Insulator}
We present an analytic theory for electron transport in a band separated from the Fermi energy by $\Delta$ with the dispersion relation
\begin{equation}
    \epsilon(p)=\Delta+\frac{p^2}{2m}.
\end{equation}
In the following analytic derivations, we employ the temporal gauge for the electric field by replacing the momentum $p$ by $p+eEt$ $(e=1)$ without the voltage slope in the potential as used in the numerical calculations~\cite{xichen}. In the time-domain, the retarded Green's function is defined as : 
\begin{equation}
    G^{R}_p(t_2,t_1) = -i \theta(t_2 - t_1) \langle \{ d_p(t_2),d^\dagger_p(t_1) \} \rangle
\end{equation}
Following the equations of motions of Green's function \cite{han2013solution}, with the damping parameter $\Gamma$ provided by the coupling to the particle reservoir, the retarded Green's function under a uniform DC electric field $E$ is given as
\begin{equation}
    G^R_p(t_2,t_1) = -i\theta(t_2-t_1)e^{-\Gamma|t_2-t_1|}\exp\left[
    -i\int^{t_2}_{t_1}\epsilon(p+Es)ds\right].
    \nonumber
\end{equation}
With the average time $T=(t_2+t_1)/2$ and the relative time $t=t_2-t_1$, we write
\begin{equation}
    G^R_p(t,T) = -i\theta(t)e^{-\Gamma|t|}\exp\left[
    -i\int^{T+t/2}_{T-t/2}\epsilon(p+Es)ds\right].
\end{equation}
We expect that the Green's functions in the steady-state become a function of the relative time only. In the clean limit ($W=0$), the frequency-dependent Green's function computed numerically in Eq.(\ref{eq:3}) would be the exact Fourier transform of the time-dependent steady-state Green's functions. Similarly, the lesser Green's function $G^{<}_p(t_2, t_1) = i \langle d^\dagger_p(t_1)d_p(t_2) \rangle$ can be expressed with the relative, and average time $(t,T)$ as~\cite{han2013solution}
\begin{equation}
    G^<_p(t,T) = \int ds_1\int ds_2 G^R_p(t_2,s_2)\Sigma_0^<(s_2-s_1)[G^R_p(t_1,s_1)]^*,
\end{equation}
with the dissipative (lesser) self-energy from the bath, $\Sigma_0^<(s)$ obtained as the Fourier transform of $\Sigma_0^<(\omega)=2i\Gamma\theta(-\omega)$ at zero temperature, as
\begin{equation}
    \Sigma_0^<(s) = 2i\Gamma \int_{-\infty}^0 e^{-i\omega s+\eta \omega}\frac{d\omega}{2\pi}=-\frac{\Gamma}{\pi(s+i\eta)}.
\end{equation}

After a lengthy but straightforward calculation, we get
\begin{eqnarray}
    G_p^<(t) & = & -\frac{\Gamma}{\pi}e^{-i\epsilon(p)t-i\frac{E^2t^3}{24m}}
    \int_{-\infty}^0 ds_1 \int_{-\infty}^0 ds_2
    \frac{e^{2\Gamma S}}{s+t+i\eta}\times\nonumber \\
    & & \exp\left[i(t+s)\epsilon(p+ES)+i\frac{E^2(s+t)^3}{24m}\right].
\end{eqnarray}
The average time $T$ dependence always appears in the combination of $p+ET$, which is the manifestation of the gauge-invariance, and we replace $p+ET$ by the gauge-covariant momentum $p$ and the above steady-state expression results. Here, $S=(s_2+s_1)/2$ is the average time and $s=s_2-s_1$ is the relative time. For the occupation number of the band at the electric field $E$, we have $n_p=-iG^<_p(t=0)$ and
\begin{eqnarray}
    n_p(E) &= &\frac{i\Gamma}{\pi}\int_{-\infty}^\infty ds\int_{-\infty}^{-|s|/2}dS\frac{e^{2\Gamma S}}{s+i\eta}\times \nonumber \\
    &\times &\exp\left[is\epsilon(p+ES)+\frac{iE^2s^3}{24m}\right]
    \label{eq:a6}.
\end{eqnarray}
In the zero field limit, the total population in the band becomes
\begin{equation}
    n_0=\int_{-\infty}^\infty n_p\frac{dp}{2\pi}=\frac{i}{4\pi^2}\int dp\int_{-\infty}^\infty ds
    \frac{e^{-\Gamma|s|+i\epsilon(p)s}}{s+i\eta},
\end{equation}
which after some contour integrals becomes
\begin{equation}
    n_0=\frac{1}{2\pi^2}\int_{-\infty}^\infty dp\tan^{-1}\left(\frac{\Gamma}{\epsilon(p)}\right)\approx \frac{\Gamma}{2\pi}\sqrt{\frac{2m}{\Delta}},
\end{equation}
in the small $\Gamma$ limit, Eq.~(\ref{n_loc}). 

For the current at a finite electric field, we compute the current similarly. The total momentum from the band becomes
\begin{eqnarray}
    \int p n_p\frac{dp}{2\pi} &= &\frac{i\Gamma}{\pi}\int\frac{dp}{2\pi}p\int_{-\infty}^\infty ds\int_{-\infty}^{-|s|/2}dS\frac{e^{2\Gamma S}}{s+i\eta}\times \nonumber \\
    &\times &\exp\left[is\epsilon(p+ES)+\frac{iE^2s^3}{24m}\right],
\end{eqnarray}
which becomes in the leading order of $E$ with the change of variable $p+ES\to p$. After performing an integral over $S$, the leading order of $E$ to the current becomes
\begin{eqnarray}
    & & \frac{i E}{8\pi^2\Gamma}\int dp\int_{-\infty}^\infty ds\frac{1+\Gamma|s|}{s+i\eta}e^{-\Gamma|s|+i\epsilon(p)s}
    \nonumber \\
    & = &\frac{E}{4\pi^2\Gamma}\int dp\left[\tan^{-1}\frac{\Gamma}{\epsilon(p)}-\frac{\Gamma\epsilon(p)}{\Gamma^2+\epsilon(p)^2}\right].
\end{eqnarray}
Taking the leading order contribution in $\Gamma$, we have the linear electric current
\begin{equation}
    J=\int\frac{p}{m}n_p\frac{dp}{2\pi}\approx \frac{\Gamma^2 E}{6\pi^2 m}\int\frac{dp}{\epsilon(p)^3}=\frac{\Gamma^2 E}{8\pi \sqrt{2m\Delta^5}},
\end{equation}
and the DC conductivity $\sigma=J/E$ as in Eq.~(\ref{eq:10})
\begin{equation}
    \sigma=\frac{\Gamma^2}{8\pi \sqrt{2m\Delta^5}}.
\end{equation}
We obtain the mobility $\mu=\sigma/n$ as
\begin{equation}
    \mu = \frac{\Gamma}{8m\Delta^2}.
\end{equation}
Interpreting this result in terms of the Drude theory $\mu=\tau/m$ with the scattering time $\tau$, we arrive at a surprising result $\tau=\Gamma/(8\Delta^2)$ in which the scattering time is proportional to the scattering rate. This is due to the fact that the band is only occasionally occupied with the probability proportional to $(\Gamma/\Delta)^2$ through hybridizing with the particle reservoir, and that the electrons accelerate from rest when they re-enter the band from the baths. Therefore, the mobility is proportional to $(\Gamma/\Delta)^2\cdot\Gamma^{-1}$, being consistent with the Drude interpretation.

\begin{figure}
\includegraphics[width=0.45\textwidth]{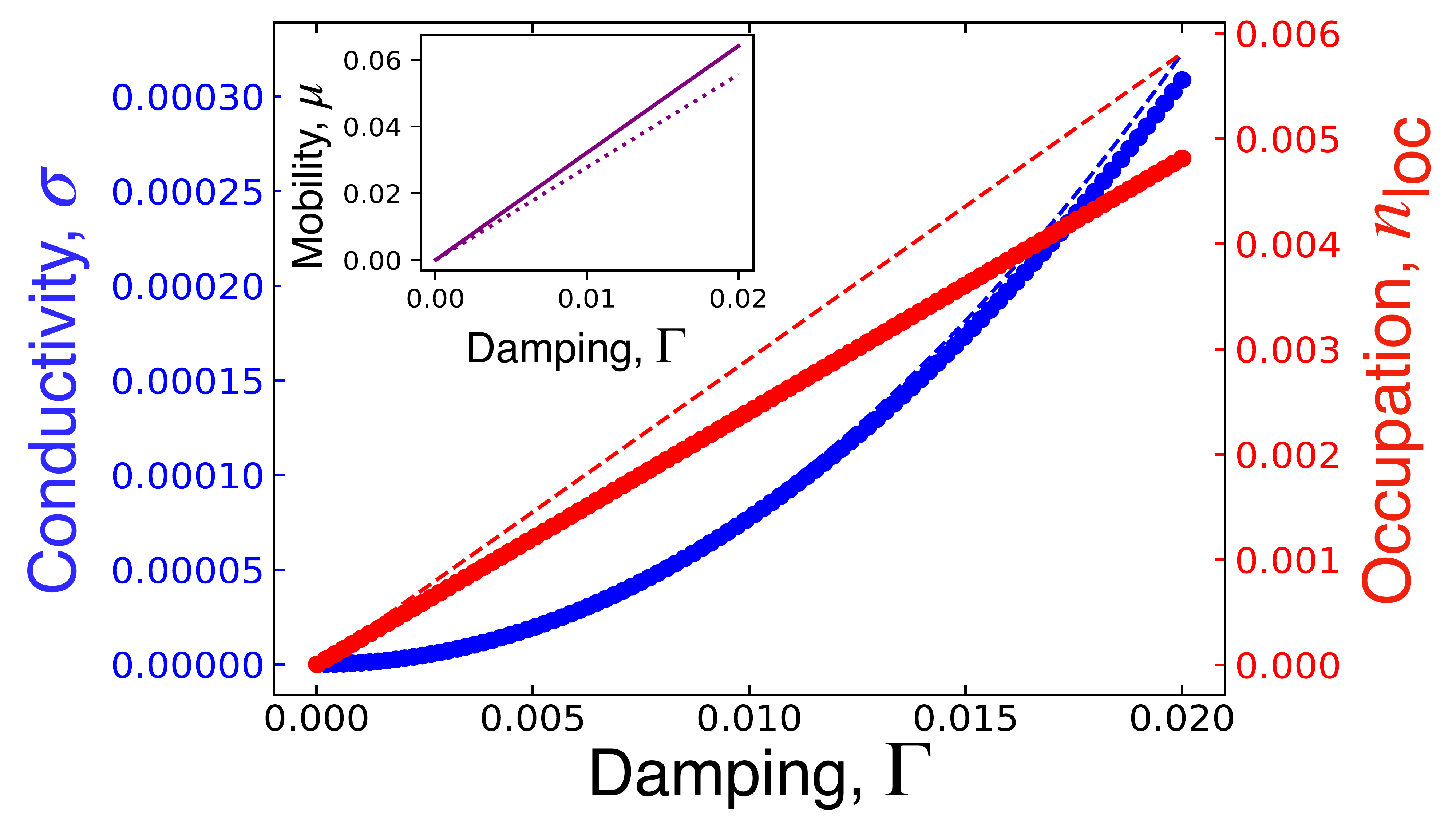}
\caption{Damping ($\Gamma$) Dependence in low temperature limit ($t =1$, $T = 0.01$, $\Delta = 0.3$) at zero electric field. The occupation varies linearly with $\Gamma$ (Red curve) and conductivity varies as $\Gamma^2$ (blue curve). The theoretical results plotted as the dashed lines seem to follow a similar trend. The inset plot shows Mobility ($\mu = \sigma/n_{\rm loc}$) varying linearly with $\Gamma$ (in purple). }
\label{fig:6}
\end{figure}

\section{ \label{App:2} Band Shift in the Coherent Potential Approximation (CPA) at Zero Field}

\begin{figure}
\centering
\includegraphics[width=\linewidth]{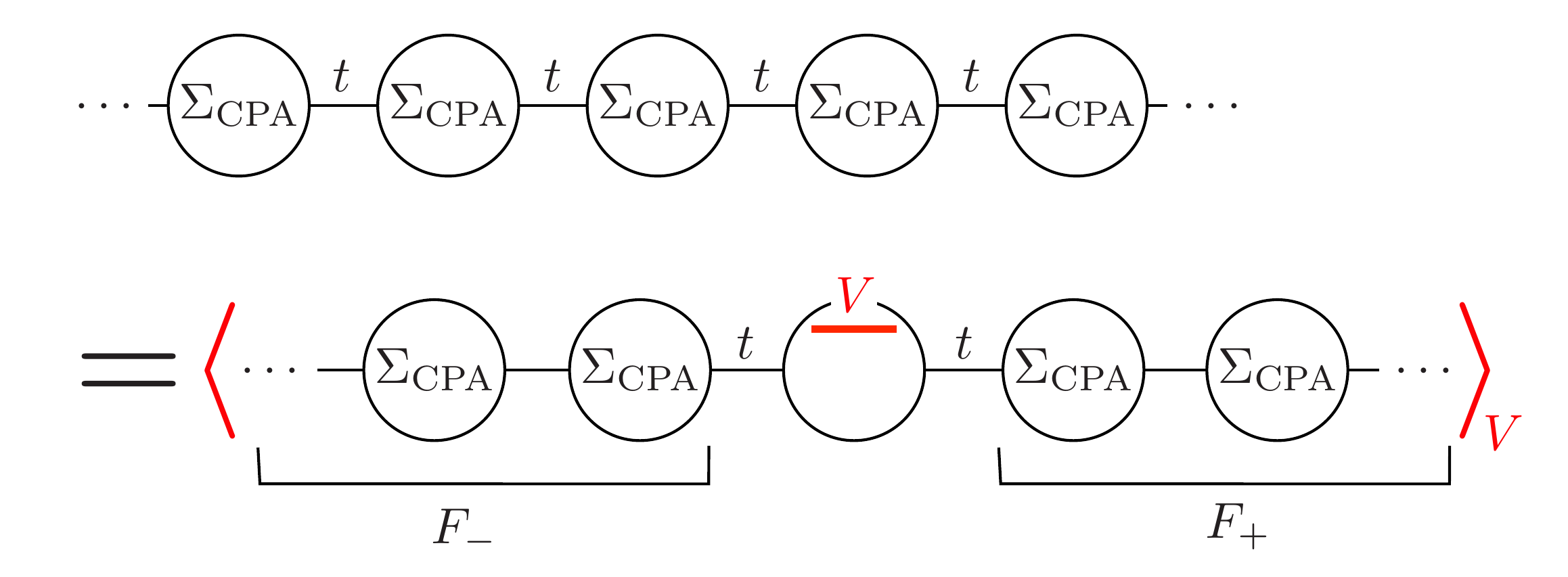}
\caption{Schematic CPA self-consistent condition. The translationally invariant lattice with the self-energy $\Sigma_{\rm CPA}$ is equivalent to the impurity-level averaging over $V$ (denoted as red) on a chosen site, with remaining sites with the embedded self-energy $\Sigma_{\rm CPA}$.}
\label{fig:cpa_scheme}
\end{figure} 

In the CPA, the effect of the disorder is represented by a self-energy $\Sigma_{\rm CPA}(\omega)$ that encodes the level shift and the dephasing from the scattering. This complex CPA self-energy can be self-consistently computed by our numerical method described in Sec-\ref{sec:level2-2}, but can be analytically verified. In the CPA method, the primary assumption is that the disordered averaged $t$-matrix becomes zero at the local impurity site \cite{PhysRev.156.809, janivs2021dynamical} and hence we can write it in the Soven equation~\cite{PhysRev.156.809} : 
\begin{equation} \label{eq:soven}
    \left\langle \frac{V - \Sigma_{\text{CPA}}(\omega)}{1 - [V - \Sigma_{\text{CPA}}(\omega)] \cdot G^R_{\text{loc}}(\omega) } \right\rangle_{V} = 0
\end{equation}
where $\left\langle f \right\rangle_{V}$ denotes the disorder averaging. We multiply the $V-$independent $G^R_{\text{loc}}(\omega)$ to both sides of Eq.~(\ref{eq:soven}) obtaining 
\begin{equation} 
    \left\langle \frac{V - \Sigma_{\text{CPA}}(\omega)}{G^R_{\text{loc}}(\omega)^{-1} - V + \Sigma_{\text{CPA}}(\omega)  } \right\rangle_{V} = 0
\end{equation}
Following the Dyson equation $G^R_{\text{loc}}(\omega) = [\mathcal{G}^{R}_{0}(\omega)^{-1} - \Sigma^{R}_{\text{CPA}}(\omega)]^{-1}$ with the DMFT Weiss-Green's function ${\cal G}_0^R(\omega)$~\cite{aoki2014nonequilibrium,georges1996dynamical} that includes its local dissipation and the contributions from the semi-infinite chains $F_{\pm}(\omega)$ but excludes the local impurity shift $V$, we can simplify the above expression as  
\begin{equation}
    \left\langle \frac{V - \Sigma_{\text{CPA}}(\omega)}{\mathcal{G}^{R}_{0}(\omega)^{-1} - V } \right\rangle_{V} = 0
\end{equation}
After some algebra, we convert the equation into Eq.~(\ref{eq:gfcpa}) as
\begin{equation} \label{eq:selfcons}
    \left\langle \frac{ 1}{\mathcal{G}^{R}_{0}(\omega)^{-1} - V} \right\rangle_V = \frac{1}{\mathcal{G}^{R}_{0}(\omega)^{-1} - \Sigma_{\text{CPA}}(\omega)} = G^R_{\text{loc}}(\omega).
\end{equation}
This is schematically depicted in Fig.~\ref{fig:cpa_scheme}, where the local Green's function is equivalent to the disordered averaged Green's function when self-consistency is achieved.

In our calculations, we consider the level-disorder where the local orbital energy is randomly shifted by $V\in[-W,W]$ set by the disorder strength $W$ and drawn from a uniform distribution. The local retarded Green's function (first line in Fig.~\ref{fig:cpa_scheme}) in the absence of an electric field in a one-dimensional chain is 
\begin{equation}
    G^R_{\rm loc}(\omega) = \int\frac{dk}{2\pi}\frac{1}{\omega-\epsilon_k-\Sigma_{\rm CPA}(\omega)},
    \label{app:gloc}
\end{equation}
with the tight-binding dispersion relation $\epsilon_k=-2t\cos k$. Here, we set the damping $\Gamma\to 0$ for convenience. The second line in FIG.~\ref{fig:cpa_scheme} represents the disorder-averaged local Green's function. The disorder-specific Green's function has the leads of semi-infinite chains with the same self-energy $\Sigma_{\rm CPA}$ and the disorder shift $V$, as
\begin{equation}
    G^R_V(\omega)=\frac{1}{\omega-V-2t^2\bar{F}^R(\omega)},
\end{equation}
where $\bar{F}^R(\omega)$ is the Green's function at the edge site that connects to the central site for each semi-infinite chain. The factor 2 is due to the sum of the left- and right-chains, $\bar{F}_-$ and $\bar{F}_+$, respectively. At zero bias, $\bar{F}_+=\bar{F}_-\equiv \bar{F}$. The $F$-Green's functions in one-dimension can be obtained recursively~\cite{li2015electric} as
\begin{equation}
    \bar{F}^R_\pm(\omega)=[\omega+i\eta -\Sigma_{\rm CPA}(\omega)-t^2 \bar{F}^R_\pm(\omega)]^{-1}.
    \label{app:F}
\end{equation}
The CPA is to equate both sides after the local averaging over the disorder $V$ is performed locally, 
\begin{equation}
G^R_{\rm loc}(\omega)=\langle G^R_V(\omega)\rangle_V
=\int^W_{-W}\frac{dV}{2W}G^R_V(\omega).
\label{app:cpa}
\end{equation}

The CPA problem, Eqs.~(\ref{app:gloc}-\ref{app:cpa}), can be solved exactly at zero bias. The $V$-integral can be performed exactly as
\begin{equation}
G^R_{\rm loc}(\omega)=\frac{1}{2W}\log\left|\frac{\omega+W-2t^2\bar{F}(\omega)}{\omega-W-2t^2\bar{F}(\omega)}\right|.
\end{equation}
Eq.~(\ref{app:gloc}) can be rewritten in the 1-$d$ chain as
\begin{equation}
    G^R_{\rm loc}(\omega)=\frac{1}{\omega-\Sigma_{\rm CPA}(\omega)-2t^2\bar{F}(\omega)}.
\end{equation}
From Eq.~(\ref{app:F}), we may eliminate $\Sigma_{\rm CPA}$ and have the self-consistent relation $f(\bar{F})=0$ with
\begin{equation}
f(\bar{F})=\frac{1}{-t^2\bar{F}+\bar{F}^{-1}}-\frac{1}{2W}\log\left|\frac{\omega+W-2t^2\bar{F}}{\omega-W-2t^2\bar{F}}\right|.
\end{equation}
By finding the root of the equation $f(\bar{F})=0$ per given set $(\omega+i\eta,W)$ we can solve the problem. To evaluate the band edge shift, we note that the spectral weight of the retarded Green's function goes to zero outside the band, \textit{i.e.} any real root $\bar{F}$ does not exist. Therefore, at the band edge we require simultaneously $f(\bar{F})=0$ and $f'(\bar{F})=0$, which sets a condition for a critical value $\omega=\omega_c$ at a given $W$. By combining the conditions, we obtain
\begin{eqnarray}
    \omega_c-2t^2\bar{F}_c & = & g(W,\bar{F}_c) 
    \label{app:wc}\\
    \frac{\bar{F}_c}{1-t^2\bar{F}_c^2} & = &
    -\frac{1}{2W}\log\left|\frac{g(W,\bar{F}_c)+W}{g(W,\bar{F}_c)-W}\right|
    \label{app:gwf}
\end{eqnarray}
with $g(W,\bar{F}_c)=[W^2+2t^2(1-t^2\bar{F}_c^2)^2/(1+t^2\bar{F}_c^2)]^{1/2}$ and $\bar{F}_c=\bar{F}(\omega_c)$. In the $W\to 0$ limit, $\omega_c=-2t$ becomes the lower band edge and the solution is $t\bar{F}_c=-1$. Therefore, for a given $W$, we solve for $\bar{F}_c$ from Eq.~(\ref{app:gwf}) in the neighborhood of $t\bar{F}_c=-1$  and obtain $\omega_c$ from Eq.~(\ref{app:wc}) for the bande edge shift $\delta$. This analytic solution agrees with the numerical CPA solution very well as shown in FIG.~\ref{fig:10}.
\nocite{*}

\bibliography{vrh1}

\providecommand{\noopsort}[1]{}\providecommand{\singleletter}[1]{#1}%
\begin{thebibliography}{57}%
\makeatletter
\providecommand \@ifxundefined [1]{%
 \@ifx{#1\undefined}
}%
\providecommand \@ifnum [1]{%
 \ifnum #1\expandafter \@firstoftwo
 \else \expandafter \@secondoftwo
 \fi
}%
\providecommand \@ifx [1]{%
 \ifx #1\expandafter \@firstoftwo
 \else \expandafter \@secondoftwo
 \fi
}%
\providecommand \natexlab [1]{#1}%
\providecommand \enquote  [1]{``#1''}%
\providecommand \bibnamefont  [1]{#1}%
\providecommand \bibfnamefont [1]{#1}%
\providecommand \citenamefont [1]{#1}%
\providecommand \href@noop [0]{\@secondoftwo}%
\providecommand \href [0]{\begingroup \@sanitize@url \@href}%
\providecommand \@href[1]{\@@startlink{#1}\@@href}%
\providecommand \@@href[1]{\endgroup#1\@@endlink}%
\providecommand \@sanitize@url [0]{\catcode `\\12\catcode `\$12\catcode
  `\&12\catcode `\#12\catcode `\^12\catcode `\_12\catcode `\%12\relax}%
\providecommand \@@startlink[1]{}%
\providecommand \@@endlink[0]{}%
\providecommand \url  [0]{\begingroup\@sanitize@url \@url }%
\providecommand \@url [1]{\endgroup\@href {#1}{\urlprefix }}%
\providecommand \urlprefix  [0]{URL }%
\providecommand \Eprint [0]{\href }%
\providecommand \doibase [0]{https://doi.org/}%
\providecommand \selectlanguage [0]{\@gobble}%
\providecommand \bibinfo  [0]{\@secondoftwo}%
\providecommand \bibfield  [0]{\@secondoftwo}%
\providecommand \translation [1]{[#1]}%
\providecommand \BibitemOpen [0]{}%
\providecommand \bibitemStop [0]{}%
\providecommand \bibitemNoStop [0]{.\EOS\space}%
\providecommand \EOS [0]{\spacefactor3000\relax}%
\providecommand \BibitemShut  [1]{\csname bibitem#1\endcsname}%
\let\auto@bib@innerbib\@empty
\bibitem [{\citenamefont {Anderson}(1958)}]{anderson1958absence}%
  \BibitemOpen
  \bibfield  {author} {\bibinfo {author} {\bibfnamefont {P.~W.}\ \bibnamefont
  {Anderson}},\ }\href {https://doi.org/10.1103/PhysRev.109.1492} {\bibfield
  {journal} {\bibinfo  {journal} {Phys. Rev.}\ }\textbf {\bibinfo {volume}
  {109}},\ \bibinfo {pages} {1492} (\bibinfo {year} {1958})}\BibitemShut
  {NoStop}%
\bibitem [{\citenamefont {Lee}\ and\ \citenamefont
  {Ramakrishnan}(1985)}]{RevModPhys.57.287}%
  \BibitemOpen
  \bibfield  {author} {\bibinfo {author} {\bibfnamefont {P.~A.}\ \bibnamefont
  {Lee}}\ and\ \bibinfo {author} {\bibfnamefont {T.~V.}\ \bibnamefont
  {Ramakrishnan}},\ }\href {https://doi.org/10.1103/RevModPhys.57.287}
  {\bibfield  {journal} {\bibinfo  {journal} {Rev. Mod. Phys.}\ }\textbf
  {\bibinfo {volume} {57}},\ \bibinfo {pages} {287} (\bibinfo {year}
  {1985})}\BibitemShut {NoStop}%
\bibitem [{\citenamefont {Mott}(1967)}]{mott1967electrons}%
  \BibitemOpen
  \bibfield  {author} {\bibinfo {author} {\bibfnamefont {N.}~\bibnamefont
  {Mott}},\ }\href {https://doi.org/10.1080/00018736700101265} {\bibfield
  {journal} {\bibinfo  {journal} {Advances in Physics}\ }\textbf {\bibinfo
  {volume} {16}},\ \bibinfo {pages} {49} (\bibinfo {year} {1967})}\BibitemShut
  {NoStop}%
\bibitem [{\citenamefont {Mott}(1968{\natexlab{a}})}]{mott1968conduction1}%
  \BibitemOpen
  \bibfield  {author} {\bibinfo {author} {\bibfnamefont {N.~F.}\ \bibnamefont
  {Mott}},\ }\href {https://doi.org/10.1080/14786436808223200} {\bibfield
  {journal} {\bibinfo  {journal} {The Phil. Mag.}\ }\textbf {\bibinfo {volume}
  {17}},\ \bibinfo {pages} {1259} (\bibinfo {year}
  {1968}{\natexlab{a}})}\BibitemShut {NoStop}%
\bibitem [{\citenamefont {Thouless}(1974)}]{thouless1974electrons}%
  \BibitemOpen
  \bibfield  {author} {\bibinfo {author} {\bibfnamefont {D.}~\bibnamefont
  {Thouless}},\ }\href
  {https://doi.org/https://doi.org/10.1016/0370-1573(74)90029-5} {\bibfield
  {journal} {\bibinfo  {journal} {Physics Reports}\ }\textbf {\bibinfo {volume}
  {13}},\ \bibinfo {pages} {93} (\bibinfo {year} {1974})}\BibitemShut {NoStop}%
\bibitem [{\citenamefont {Kramer}\ and\ \citenamefont
  {MacKinnon}(1993)}]{kramer1993localization}%
  \BibitemOpen
  \bibfield  {author} {\bibinfo {author} {\bibfnamefont {B.}~\bibnamefont
  {Kramer}}\ and\ \bibinfo {author} {\bibfnamefont {A.}~\bibnamefont
  {MacKinnon}},\ }\href {https://doi.org/10.1088/0034-4885/56/12/001}
  {\bibfield  {journal} {\bibinfo  {journal} {Rep. Prog. Phys.}\ }\textbf
  {\bibinfo {volume} {56}},\ \bibinfo {pages} {1469} (\bibinfo {year}
  {1993})}\BibitemShut {NoStop}%
\bibitem [{\citenamefont {Anderson}(2010)}]{anderson201050}%
  \BibitemOpen
  \bibfield  {author} {\bibinfo {author} {\bibfnamefont {P.}~\bibnamefont
  {Anderson}},\ }\bibfield  {title} {\bibinfo {title} {\textit{50 Years of
  Anderson Localization}},\ }\href@noop {} {\bibfield  {journal} {\bibinfo
  {journal} {Published by World Scientific Publishing Co. Pte. Ltd}\ }
  (\bibinfo {year} {2010})}\BibitemShut {NoStop}%
\bibitem [{\citenamefont {Cutler}\ and\ \citenamefont
  {Mott}(1969)}]{cutler1969observation}%
  \BibitemOpen
  \bibfield  {author} {\bibinfo {author} {\bibfnamefont {M.}~\bibnamefont
  {Cutler}}\ and\ \bibinfo {author} {\bibfnamefont {N.~F.}\ \bibnamefont
  {Mott}},\ }\href {https://doi.org/10.1103/PhysRev.181.1336} {\bibfield
  {journal} {\bibinfo  {journal} {Phys. Rev.}\ }\textbf {\bibinfo {volume}
  {181}},\ \bibinfo {pages} {1336} (\bibinfo {year} {1969})}\BibitemShut
  {NoStop}%
\bibitem [{\citenamefont {Sheng}(1990)}]{sheng1990scattering}%
  \BibitemOpen
  \bibfield  {author} {\bibinfo {author} {\bibfnamefont {P.}~\bibnamefont
  {Sheng}},\ }\href@noop {} {\emph {\bibinfo {title} {Scattering and
  localization of classical waves in random media}}},\ Vol.~\bibinfo {volume}
  {8}\ (\bibinfo  {publisher} {World Scientific},\ \bibinfo {year}
  {1990})\BibitemShut {NoStop}%
\bibitem [{\citenamefont {Kirkpatrick}(1985)}]{kirkpatrick1985localization}%
  \BibitemOpen
  \bibfield  {author} {\bibinfo {author} {\bibfnamefont {T.~R.}\ \bibnamefont
  {Kirkpatrick}},\ }\href {https://doi.org/10.1103/PhysRevB.31.5746} {\bibfield
   {journal} {\bibinfo  {journal} {Phys. Rev. B}\ }\textbf {\bibinfo {volume}
  {31}},\ \bibinfo {pages} {5746} (\bibinfo {year} {1985})}\BibitemShut
  {NoStop}%
\bibitem [{\citenamefont {John}(1984)}]{john1984electromagnetic}%
  \BibitemOpen
  \bibfield  {author} {\bibinfo {author} {\bibfnamefont {S.}~\bibnamefont
  {John}},\ }\href {https://doi.org/10.1103/PhysRevLett.53.2169} {\bibfield
  {journal} {\bibinfo  {journal} {Phys. Rev. Lett.}\ }\textbf {\bibinfo
  {volume} {53}},\ \bibinfo {pages} {2169} (\bibinfo {year}
  {1984})}\BibitemShut {NoStop}%
\bibitem [{\citenamefont {Lagendijk}\ \emph {et~al.}(2009)\citenamefont
  {Lagendijk}, \citenamefont {Tiggelen},\ and\ \citenamefont
  {Wiersma}}]{lagendijk2009fifty}%
  \BibitemOpen
  \bibfield  {author} {\bibinfo {author} {\bibfnamefont {A.}~\bibnamefont
  {Lagendijk}}, \bibinfo {author} {\bibfnamefont {B.~v.}\ \bibnamefont
  {Tiggelen}},\ and\ \bibinfo {author} {\bibfnamefont {D.~S.}\ \bibnamefont
  {Wiersma}},\ }\href@noop {} {\bibfield  {journal} {\bibinfo  {journal}
  {Physics Today}\ }\textbf {\bibinfo {volume} {62}},\ \bibinfo {pages} {24}
  (\bibinfo {year} {2009})}\BibitemShut {NoStop}%
\bibitem [{\citenamefont {Wiersma}\ \emph {et~al.}(1997)\citenamefont
  {Wiersma}, \citenamefont {Bartolini}, \citenamefont {Lagendijk},\ and\
  \citenamefont {Righini}}]{wiersma1997localization}%
  \BibitemOpen
  \bibfield  {author} {\bibinfo {author} {\bibfnamefont {D.~S.}\ \bibnamefont
  {Wiersma}}, \bibinfo {author} {\bibfnamefont {P.}~\bibnamefont {Bartolini}},
  \bibinfo {author} {\bibfnamefont {A.}~\bibnamefont {Lagendijk}},\ and\
  \bibinfo {author} {\bibfnamefont {R.}~\bibnamefont {Righini}},\ }\href
  {https://doi.org/10.1038/37757} {\bibfield  {journal} {\bibinfo  {journal}
  {Nature}\ }\textbf {\bibinfo {volume} {390}},\ \bibinfo {pages} {671}
  (\bibinfo {year} {1997})}\BibitemShut {NoStop}%
\bibitem [{\citenamefont {Schwartz}\ \emph {et~al.}(2007)\citenamefont
  {Schwartz}, \citenamefont {Bartal}, \citenamefont {Fishman},\ and\
  \citenamefont {Segev}}]{schwartz2007transport}%
  \BibitemOpen
  \bibfield  {author} {\bibinfo {author} {\bibfnamefont {T.}~\bibnamefont
  {Schwartz}}, \bibinfo {author} {\bibfnamefont {G.}~\bibnamefont {Bartal}},
  \bibinfo {author} {\bibfnamefont {S.}~\bibnamefont {Fishman}},\ and\ \bibinfo
  {author} {\bibfnamefont {M.}~\bibnamefont {Segev}},\ }\href
  {https://doi.org/10.1038/nature05623} {\bibfield  {journal} {\bibinfo
  {journal} {Nature}\ }\textbf {\bibinfo {volume} {446}},\ \bibinfo {pages}
  {52} (\bibinfo {year} {2007})}\BibitemShut {NoStop}%
\bibitem [{\citenamefont {Segev}\ \emph {et~al.}(2013)\citenamefont {Segev},
  \citenamefont {Silberberg},\ and\ \citenamefont
  {Christodoulides}}]{segev2013anderson}%
  \BibitemOpen
  \bibfield  {author} {\bibinfo {author} {\bibfnamefont {M.}~\bibnamefont
  {Segev}}, \bibinfo {author} {\bibfnamefont {Y.}~\bibnamefont {Silberberg}},\
  and\ \bibinfo {author} {\bibfnamefont {D.~N.}\ \bibnamefont
  {Christodoulides}},\ }\href {https://doi.org/10.1038/nphoton.2013.30}
  {\bibfield  {journal} {\bibinfo  {journal} {Nat. Photonics}\ }\textbf
  {\bibinfo {volume} {7}},\ \bibinfo {pages} {197} (\bibinfo {year}
  {2013})}\BibitemShut {NoStop}%
\bibitem [{\citenamefont {Rothstein}(2013)}]{rothstein2013gravitational}%
  \BibitemOpen
  \bibfield  {author} {\bibinfo {author} {\bibfnamefont {I.~Z.}\ \bibnamefont
  {Rothstein}},\ }\href {https://doi.org/10.1103/PhysRevLett.110.011601}
  {\bibfield  {journal} {\bibinfo  {journal} {Phys. Rev. Lett.}\ }\textbf
  {\bibinfo {volume} {110}},\ \bibinfo {pages} {011601} (\bibinfo {year}
  {2013})}\BibitemShut {NoStop}%
\bibitem [{\citenamefont {Tian}(2019)}]{tian2019anderson}%
  \BibitemOpen
  \bibfield  {author} {\bibinfo {author} {\bibfnamefont {Z.}~\bibnamefont
  {Tian}},\ }\href {https://doi.org/10.1021/acsnano.9b02399} {\bibfield
  {journal} {\bibinfo  {journal} {ACS Nano}\ }\textbf {\bibinfo {volume}
  {13}},\ \bibinfo {pages} {3750} (\bibinfo {year} {2019})}\BibitemShut
  {NoStop}%
\bibitem [{\citenamefont {Mafi}\ \emph {et~al.}(2019)\citenamefont {Mafi},
  \citenamefont {Ballato}, \citenamefont {Koch},\ and\ \citenamefont
  {Schülzgen}}]{mafi2019disordered}%
  \BibitemOpen
  \bibfield  {author} {\bibinfo {author} {\bibfnamefont {A.}~\bibnamefont
  {Mafi}}, \bibinfo {author} {\bibfnamefont {J.}~\bibnamefont {Ballato}},
  \bibinfo {author} {\bibfnamefont {K.~W.}\ \bibnamefont {Koch}},\ and\
  \bibinfo {author} {\bibfnamefont {A.}~\bibnamefont {Schülzgen}},\ }\href
  {https://doi.org/10.1109/JLT.2019.2916020} {\bibfield  {journal} {\bibinfo
  {journal} {J Lightwave Technology}\ }\textbf {\bibinfo {volume} {37}},\
  \bibinfo {pages} {5652} (\bibinfo {year} {2019})}\BibitemShut {NoStop}%
\bibitem [{\citenamefont {Mott}\ \emph {et~al.}(1975)\citenamefont {Mott},
  \citenamefont {Pepper}, \citenamefont {Pollitt}, \citenamefont {Wallis},\
  and\ \citenamefont {Adkins}}]{mott1975anderson}%
  \BibitemOpen
  \bibfield  {author} {\bibinfo {author} {\bibfnamefont {N.~F.}\ \bibnamefont
  {Mott}}, \bibinfo {author} {\bibfnamefont {M.}~\bibnamefont {Pepper}},
  \bibinfo {author} {\bibfnamefont {S.}~\bibnamefont {Pollitt}}, \bibinfo
  {author} {\bibfnamefont {R.~H.}\ \bibnamefont {Wallis}},\ and\ \bibinfo
  {author} {\bibfnamefont {C.~J.}\ \bibnamefont {Adkins}},\ }\href
  {https://doi.org/10.1098/rspa.1975.0131} {\bibfield  {journal} {\bibinfo
  {journal} {Proc. of the Royal Soc. of London. A. Math. and Phys. Sci.}\
  }\textbf {\bibinfo {volume} {345}},\ \bibinfo {pages} {169} (\bibinfo {year}
  {1975})}\BibitemShut {NoStop}%
\bibitem [{\citenamefont {Mott}(1971)}]{mott1971conduction}%
  \BibitemOpen
  \bibfield  {author} {\bibinfo {author} {\bibfnamefont {N.~F.}\ \bibnamefont
  {Mott}},\ }\href {https://doi.org/10.1080/14786437108217058} {\bibfield
  {journal} {\bibinfo  {journal} {Phil. Mag.}\ }\textbf {\bibinfo {volume}
  {24}},\ \bibinfo {pages} {911} (\bibinfo {year} {1971})}\BibitemShut
  {NoStop}%
\bibitem [{\citenamefont {Mott}\ and\ \citenamefont
  {Twose}(1961)}]{Mott01041961}%
  \BibitemOpen
  \bibfield  {author} {\bibinfo {author} {\bibfnamefont {N.}~\bibnamefont
  {Mott}}\ and\ \bibinfo {author} {\bibfnamefont {W.}~\bibnamefont {Twose}},\
  }\href {https://doi.org/10.1080/00018736100101271} {\bibfield  {journal}
  {\bibinfo  {journal} {Advances in Physics}\ }\textbf {\bibinfo {volume}
  {10}},\ \bibinfo {pages} {107} (\bibinfo {year} {1961})}\BibitemShut
  {NoStop}%
\bibitem [{\citenamefont {Halperin}(1965)}]{halperin1965green}%
  \BibitemOpen
  \bibfield  {author} {\bibinfo {author} {\bibfnamefont {B.~I.}\ \bibnamefont
  {Halperin}},\ }\href {https://doi.org/10.1103/PhysRev.139.A104} {\bibfield
  {journal} {\bibinfo  {journal} {Phys. Rev.}\ }\textbf {\bibinfo {volume}
  {139}},\ \bibinfo {pages} {A104} (\bibinfo {year} {1965})}\BibitemShut
  {NoStop}%
\bibitem [{\citenamefont {Garcia}\ and\ \citenamefont
  {Hofmann}(2024)}]{garcia2024fluctuation}%
  \BibitemOpen
  \bibfield  {author} {\bibinfo {author} {\bibfnamefont {E.~R.}\ \bibnamefont
  {Garcia}}\ and\ \bibinfo {author} {\bibfnamefont {J.}~\bibnamefont
  {Hofmann}},\ }\href {https://doi.org/10.1103/PhysRevE.109.L032103} {\bibfield
   {journal} {\bibinfo  {journal} {Phys. Rev. E}\ }\textbf {\bibinfo {volume}
  {109}},\ \bibinfo {pages} {L032103} (\bibinfo {year} {2024})}\BibitemShut
  {NoStop}%
\bibitem [{\citenamefont {Van~Mieghem}(1992)}]{RevModPhys.64.755}%
  \BibitemOpen
  \bibfield  {author} {\bibinfo {author} {\bibfnamefont {P.}~\bibnamefont
  {Van~Mieghem}},\ }\href {https://doi.org/10.1103/RevModPhys.64.755}
  {\bibfield  {journal} {\bibinfo  {journal} {Rev. Mod. Phys.}\ }\textbf
  {\bibinfo {volume} {64}},\ \bibinfo {pages} {755} (\bibinfo {year}
  {1992})}\BibitemShut {NoStop}%
\bibitem [{\citenamefont {Lee}\ \emph {et~al.}(1983)\citenamefont {Lee},
  \citenamefont {Chu},\ and\ \citenamefont {Casta\~no}}]{lee1983effects}%
  \BibitemOpen
  \bibfield  {author} {\bibinfo {author} {\bibfnamefont {Y.~C.}\ \bibnamefont
  {Lee}}, \bibinfo {author} {\bibfnamefont {C.~S.}\ \bibnamefont {Chu}},\ and\
  \bibinfo {author} {\bibfnamefont {E.}~\bibnamefont {Casta\~no}},\ }\href
  {https://doi.org/10.1103/PhysRevB.27.6136} {\bibfield  {journal} {\bibinfo
  {journal} {Phys. Rev. B}\ }\textbf {\bibinfo {volume} {27}},\ \bibinfo
  {pages} {6136} (\bibinfo {year} {1983})}\BibitemShut {NoStop}%
\bibitem [{\citenamefont {Soukoulis}\ \emph {et~al.}(1983)\citenamefont
  {Soukoulis}, \citenamefont {Jos\'e}, \citenamefont {Economou},\ and\
  \citenamefont {Sheng}}]{soukoulis1983localization}%
  \BibitemOpen
  \bibfield  {author} {\bibinfo {author} {\bibfnamefont {C.~M.}\ \bibnamefont
  {Soukoulis}}, \bibinfo {author} {\bibfnamefont {J.~V.}\ \bibnamefont
  {Jos\'e}}, \bibinfo {author} {\bibfnamefont {E.~N.}\ \bibnamefont
  {Economou}},\ and\ \bibinfo {author} {\bibfnamefont {P.}~\bibnamefont
  {Sheng}},\ }\href {https://doi.org/10.1103/PhysRevLett.50.764} {\bibfield
  {journal} {\bibinfo  {journal} {Phys. Rev. Lett.}\ }\textbf {\bibinfo
  {volume} {50}},\ \bibinfo {pages} {764} (\bibinfo {year} {1983})}\BibitemShut
  {NoStop}%
\bibitem [{\citenamefont {Prigodin}(1980)}]{prigodin1980one}%
  \BibitemOpen
  \bibfield  {author} {\bibinfo {author} {\bibfnamefont {V.}~\bibnamefont
  {Prigodin}},\ }\href@noop {} {\bibfield  {journal} {\bibinfo  {journal} {Zh.
  Eksp. Teor. Fiz}\ }\textbf {\bibinfo {volume} {79}},\ \bibinfo {pages} {2338}
  (\bibinfo {year} {1980})}\BibitemShut {NoStop}%
\bibitem [{\citenamefont {Kirkpatrick}(1986)}]{kirkpatrick1986anderson}%
  \BibitemOpen
  \bibfield  {author} {\bibinfo {author} {\bibfnamefont {T.~R.}\ \bibnamefont
  {Kirkpatrick}},\ }\href {https://doi.org/10.1103/PhysRevB.33.780} {\bibfield
  {journal} {\bibinfo  {journal} {Phys. Rev. B}\ }\textbf {\bibinfo {volume}
  {33}},\ \bibinfo {pages} {780} (\bibinfo {year} {1986})}\BibitemShut
  {NoStop}%
\bibitem [{\citenamefont {Prigodin}\ and\ \citenamefont
  {Altshuler}(1989)}]{prigodin1989localization}%
  \BibitemOpen
  \bibfield  {author} {\bibinfo {author} {\bibfnamefont {V.~N.}\ \bibnamefont
  {Prigodin}}\ and\ \bibinfo {author} {\bibfnamefont {B.~L.}\ \bibnamefont
  {Altshuler}},\ }\href {https://api.semanticscholar.org/CorpusID:121790172}
  {\bibfield  {journal} {\bibinfo  {journal} {Phys. Lett. A}\ }\textbf
  {\bibinfo {volume} {137}},\ \bibinfo {pages} {301} (\bibinfo {year}
  {1989})}\BibitemShut {NoStop}%
\bibitem [{\citenamefont {Bleibaum}\ \emph {et~al.}(1995)\citenamefont
  {Bleibaum}, \citenamefont {B\"ottger}, \citenamefont {Bryksin},\ and\
  \citenamefont {Kleinert}}]{bleibaum1995theory}%
  \BibitemOpen
  \bibfield  {author} {\bibinfo {author} {\bibfnamefont {O.}~\bibnamefont
  {Bleibaum}}, \bibinfo {author} {\bibfnamefont {H.}~\bibnamefont {B\"ottger}},
  \bibinfo {author} {\bibfnamefont {V.~V.}\ \bibnamefont {Bryksin}},\ and\
  \bibinfo {author} {\bibfnamefont {P.}~\bibnamefont {Kleinert}},\ }\href
  {https://doi.org/10.1103/PhysRevB.52.16494} {\bibfield  {journal} {\bibinfo
  {journal} {Phys. Rev. B}\ }\textbf {\bibinfo {volume} {52}},\ \bibinfo
  {pages} {16494} (\bibinfo {year} {1995})}\BibitemShut {NoStop}%
\bibitem [{\citenamefont {Bleibaum}\ and\ \citenamefont
  {Belitz}(2004)}]{bleibaum2004weak}%
  \BibitemOpen
  \bibfield  {author} {\bibinfo {author} {\bibfnamefont {O.}~\bibnamefont
  {Bleibaum}}\ and\ \bibinfo {author} {\bibfnamefont {D.}~\bibnamefont
  {Belitz}},\ }\href {https://doi.org/10.1103/PhysRevB.69.075119} {\bibfield
  {journal} {\bibinfo  {journal} {Phys. Rev. B}\ }\textbf {\bibinfo {volume}
  {69}},\ \bibinfo {pages} {075119} (\bibinfo {year} {2004})}\BibitemShut
  {NoStop}%
\bibitem [{\citenamefont {Mott}(1968{\natexlab{b}})}]{mott1968conduction}%
  \BibitemOpen
  \bibfield  {author} {\bibinfo {author} {\bibfnamefont {N.~F.}\ \bibnamefont
  {Mott}},\ }\href
  {https://www.sciencedirect.com/science/article/pii/0022309368900021}
  {\bibfield  {journal} {\bibinfo  {journal} {J. Non-Cryst. Solids}\ }\textbf
  {\bibinfo {volume} {1}},\ \bibinfo {pages} {1} (\bibinfo {year}
  {1968}{\natexlab{b}})}\BibitemShut {NoStop}%
\bibitem [{\citenamefont {Miller}\ and\ \citenamefont
  {Abrahams}(1960)}]{PhysRev.120.745}%
  \BibitemOpen
  \bibfield  {author} {\bibinfo {author} {\bibfnamefont {A.}~\bibnamefont
  {Miller}}\ and\ \bibinfo {author} {\bibfnamefont {E.}~\bibnamefont
  {Abrahams}},\ }\href {https://doi.org/10.1103/PhysRev.120.745} {\bibfield
  {journal} {\bibinfo  {journal} {Phys. Rev.}\ }\textbf {\bibinfo {volume}
  {120}},\ \bibinfo {pages} {745} (\bibinfo {year} {1960})}\BibitemShut
  {NoStop}%
\bibitem [{\citenamefont {Shklovskii}(2024)}]{10.1063/10.0034343}%
  \BibitemOpen
  \bibfield  {author} {\bibinfo {author} {\bibfnamefont {B.~I.}\ \bibnamefont
  {Shklovskii}},\ }\href {https://doi.org/10.1063/10.0034343} {\bibfield
  {journal} {\bibinfo  {journal} {Low Temp. Phys.}\ }\textbf {\bibinfo {volume}
  {50}},\ \bibinfo {pages} {1101} (\bibinfo {year} {2024})}\BibitemShut
  {NoStop}%
\bibitem [{\citenamefont {Lee}(1984)}]{lee1984variable}%
  \BibitemOpen
  \bibfield  {author} {\bibinfo {author} {\bibfnamefont {P.~A.}\ \bibnamefont
  {Lee}},\ }\href {https://doi.org/10.1103/PhysRevLett.53.2042} {\bibfield
  {journal} {\bibinfo  {journal} {Phys. Rev. Lett.}\ }\textbf {\bibinfo
  {volume} {53}},\ \bibinfo {pages} {2042} (\bibinfo {year}
  {1984})}\BibitemShut {NoStop}%
\bibitem [{\citenamefont {Shklovskii}\ and\ \citenamefont
  {Efros}(2013)}]{shklovskii2013electronic}%
  \BibitemOpen
  \bibfield  {author} {\bibinfo {author} {\bibfnamefont {B.~I.}\ \bibnamefont
  {Shklovskii}}\ and\ \bibinfo {author} {\bibfnamefont {A.~L.}\ \bibnamefont
  {Efros}},\ }\href@noop {} {\emph {\bibinfo {title} {Electronic properties of
  doped semiconductors}}},\ Vol.~\bibinfo {volume} {45}\ (\bibinfo  {publisher}
  {Springer Science \& Business Media},\ \bibinfo {year} {2013})\BibitemShut
  {NoStop}%
\bibitem [{\citenamefont {Epstein}\ \emph {et~al.}(2001)\citenamefont
  {Epstein}, \citenamefont {Lee},\ and\ \citenamefont
  {Prigodin}}]{EPSTEIN20019}%
  \BibitemOpen
  \bibfield  {author} {\bibinfo {author} {\bibfnamefont {A.}~\bibnamefont
  {Epstein}}, \bibinfo {author} {\bibfnamefont {W.-P.}\ \bibnamefont {Lee}},\
  and\ \bibinfo {author} {\bibfnamefont {V.}~\bibnamefont {Prigodin}},\ }\href
  {https://doi.org/https://doi.org/10.1016/S0379-6779(00)00531-2} {\bibfield
  {journal} {\bibinfo  {journal} {Syn. Met.}\ }\textbf {\bibinfo {volume}
  {117}},\ \bibinfo {pages} {9} (\bibinfo {year} {2001})}\BibitemShut {NoStop}%
\bibitem [{\citenamefont {Apsley}\ and\ \citenamefont
  {Hughes}(1974)}]{Apsley1974-APSTFO}%
  \BibitemOpen
  \bibfield  {author} {\bibinfo {author} {\bibfnamefont {N.}~\bibnamefont
  {Apsley}}\ and\ \bibinfo {author} {\bibfnamefont {H.~P.}\ \bibnamefont
  {Hughes}},\ }\href {https://doi.org/10.1080/14786437408207250} {\bibfield
  {journal} {\bibinfo  {journal} {Philosophical Magazine}\ }\textbf {\bibinfo
  {volume} {30}},\ \bibinfo {pages} {963} (\bibinfo {year} {1974})}\BibitemShut
  {NoStop}%
\bibitem [{\citenamefont {Liu}\ \emph {et~al.}(2010)\citenamefont {Liu},
  \citenamefont {Pourret},\ and\ \citenamefont
  {Guyot-Sionnest}}]{doi:10.1021/nn101376u}%
  \BibitemOpen
  \bibfield  {author} {\bibinfo {author} {\bibfnamefont {H.}~\bibnamefont
  {Liu}}, \bibinfo {author} {\bibfnamefont {A.}~\bibnamefont {Pourret}},\ and\
  \bibinfo {author} {\bibfnamefont {P.}~\bibnamefont {Guyot-Sionnest}},\ }\href
  {https://doi.org/10.1021/nn101376u} {\bibfield  {journal} {\bibinfo
  {journal} {ACS Nano}\ }\textbf {\bibinfo {volume} {4}},\ \bibinfo {pages}
  {5211} (\bibinfo {year} {2010})}\BibitemShut {NoStop}%
\bibitem [{\citenamefont {Paasch}\ \emph {et~al.}(2002)\citenamefont {Paasch},
  \citenamefont {Lindner},\ and\ \citenamefont {Scheinert}}]{PAASCH200297}%
  \BibitemOpen
  \bibfield  {author} {\bibinfo {author} {\bibfnamefont {G.}~\bibnamefont
  {Paasch}}, \bibinfo {author} {\bibfnamefont {T.}~\bibnamefont {Lindner}},\
  and\ \bibinfo {author} {\bibfnamefont {S.}~\bibnamefont {Scheinert}},\ }\href
  {https://doi.org/https://doi.org/10.1016/S0379-6779(02)00236-9} {\bibfield
  {journal} {\bibinfo  {journal} {Synthetic Metals}\ }\textbf {\bibinfo
  {volume} {132}},\ \bibinfo {pages} {97} (\bibinfo {year} {2002})}\BibitemShut
  {NoStop}%
\bibitem [{\citenamefont {Evans}\ \emph {et~al.}(2017)\citenamefont {Evans},
  \citenamefont {Labram}, \citenamefont {Smock}, \citenamefont {Wu},
  \citenamefont {Chabinyc}, \citenamefont {Seshadri},\ and\ \citenamefont
  {Wudl}}]{evans2017mono}%
  \BibitemOpen
  \bibfield  {author} {\bibinfo {author} {\bibfnamefont {H.~A.}\ \bibnamefont
  {Evans}}, \bibinfo {author} {\bibfnamefont {J.~G.}\ \bibnamefont {Labram}},
  \bibinfo {author} {\bibfnamefont {S.~R.}\ \bibnamefont {Smock}}, \bibinfo
  {author} {\bibfnamefont {G.}~\bibnamefont {Wu}}, \bibinfo {author}
  {\bibfnamefont {M.~L.}\ \bibnamefont {Chabinyc}}, \bibinfo {author}
  {\bibfnamefont {R.}~\bibnamefont {Seshadri}},\ and\ \bibinfo {author}
  {\bibfnamefont {F.}~\bibnamefont {Wudl}},\ }\href
  {https://doi.org/10.1021/acs.inorgchem.6b02287} {\bibfield  {journal}
  {\bibinfo  {journal} {Inorg. Chem.}\ }\textbf {\bibinfo {volume} {56}},\
  \bibinfo {pages} {395} (\bibinfo {year} {2017})}\BibitemShut {NoStop}%
\bibitem [{\citenamefont {Elliott}\ \emph {et~al.}(1974)\citenamefont
  {Elliott}, \citenamefont {Krumhansl},\ and\ \citenamefont
  {Leath}}]{elliott1974theory}%
  \BibitemOpen
  \bibfield  {author} {\bibinfo {author} {\bibfnamefont {R.~J.}\ \bibnamefont
  {Elliott}}, \bibinfo {author} {\bibfnamefont {J.~A.}\ \bibnamefont
  {Krumhansl}},\ and\ \bibinfo {author} {\bibfnamefont {P.~L.}\ \bibnamefont
  {Leath}},\ }\href {https://doi.org/10.1103/RevModPhys.46.465} {\bibfield
  {journal} {\bibinfo  {journal} {Rev. Mod. Phys.}\ }\textbf {\bibinfo {volume}
  {46}},\ \bibinfo {pages} {465} (\bibinfo {year} {1974})}\BibitemShut
  {NoStop}%
\bibitem [{\citenamefont {Jani{\v{s}}}(2021)}]{janivs2021dynamical}%
  \BibitemOpen
  \bibfield  {author} {\bibinfo {author} {\bibfnamefont {V.}~\bibnamefont
  {Jani{\v{s}}}},\ }\href@noop {} {\bibfield  {journal} {\bibinfo  {journal}
  {arXiv preprint arXiv:2109.04723}\ } (\bibinfo {year} {2021})}\BibitemShut
  {NoStop}%
\bibitem [{\citenamefont {Dohner}\ \emph {et~al.}(2022)\citenamefont {Dohner},
  \citenamefont {Terletska}, \citenamefont {Tam}, \citenamefont {Moreno},\ and\
  \citenamefont {Fotso}}]{dohner2022nonequilibrium}%
  \BibitemOpen
  \bibfield  {author} {\bibinfo {author} {\bibfnamefont {E.}~\bibnamefont
  {Dohner}}, \bibinfo {author} {\bibfnamefont {H.}~\bibnamefont {Terletska}},
  \bibinfo {author} {\bibfnamefont {K.-M.}\ \bibnamefont {Tam}}, \bibinfo
  {author} {\bibfnamefont {J.}~\bibnamefont {Moreno}},\ and\ \bibinfo {author}
  {\bibfnamefont {H.~F.}\ \bibnamefont {Fotso}},\ }\href
  {https://doi.org/10.1103/PhysRevB.106.195156} {\bibfield  {journal} {\bibinfo
   {journal} {Phys. Rev. B}\ }\textbf {\bibinfo {volume} {106}},\ \bibinfo
  {pages} {195156} (\bibinfo {year} {2022})}\BibitemShut {NoStop}%
\bibitem [{\citenamefont {Han}(2013)}]{han2013solution}%
  \BibitemOpen
  \bibfield  {author} {\bibinfo {author} {\bibfnamefont {J.~E.}\ \bibnamefont
  {Han}},\ }\href {https://doi.org/10.1103/PhysRevB.87.085119} {\bibfield
  {journal} {\bibinfo  {journal} {Phys. Rev. B}\ }\textbf {\bibinfo {volume}
  {87}},\ \bibinfo {pages} {085119} (\bibinfo {year} {2013})}\BibitemShut
  {NoStop}%
\bibitem [{\citenamefont {Li}\ \emph {et~al.}(2015)\citenamefont {Li},
  \citenamefont {Aron}, \citenamefont {Kotliar},\ and\ \citenamefont
  {Han}}]{li2015electric}%
  \BibitemOpen
  \bibfield  {author} {\bibinfo {author} {\bibfnamefont {J.}~\bibnamefont
  {Li}}, \bibinfo {author} {\bibfnamefont {C.}~\bibnamefont {Aron}}, \bibinfo
  {author} {\bibfnamefont {G.}~\bibnamefont {Kotliar}},\ and\ \bibinfo {author}
  {\bibfnamefont {J.~E.}\ \bibnamefont {Han}},\ }\href
  {https://doi.org/10.1103/PhysRevLett.114.226403} {\bibfield  {journal}
  {\bibinfo  {journal} {Phys. Rev. Lett.}\ }\textbf {\bibinfo {volume} {114}},\
  \bibinfo {pages} {226403} (\bibinfo {year} {2015})}\BibitemShut {NoStop}%
\bibitem [{\citenamefont {Haug}\ and\ \citenamefont
  {Jauho}(2008)}]{haug2008quantum}%
  \BibitemOpen
  \bibfield  {author} {\bibinfo {author} {\bibfnamefont {H.}~\bibnamefont
  {Haug}}\ and\ \bibinfo {author} {\bibfnamefont {A.-P.}\ \bibnamefont
  {Jauho}},\ }\href@noop {} {\emph {\bibinfo {title} {Quantum kinetics in
  transport and optics of semiconductors}}}\ (\bibinfo  {publisher}
  {Springer},\ \bibinfo {year} {2008})\BibitemShut {NoStop}%
\bibitem [{\citenamefont {Aoki}\ \emph {et~al.}(2014)\citenamefont {Aoki},
  \citenamefont {Tsuji}, \citenamefont {Eckstein}, \citenamefont {Kollar},
  \citenamefont {Oka},\ and\ \citenamefont {Werner}}]{aoki2014nonequilibrium}%
  \BibitemOpen
  \bibfield  {author} {\bibinfo {author} {\bibfnamefont {H.}~\bibnamefont
  {Aoki}}, \bibinfo {author} {\bibfnamefont {N.}~\bibnamefont {Tsuji}},
  \bibinfo {author} {\bibfnamefont {M.}~\bibnamefont {Eckstein}}, \bibinfo
  {author} {\bibfnamefont {M.}~\bibnamefont {Kollar}}, \bibinfo {author}
  {\bibfnamefont {T.}~\bibnamefont {Oka}},\ and\ \bibinfo {author}
  {\bibfnamefont {P.}~\bibnamefont {Werner}},\ }\href
  {https://doi.org/10.1103/RevModPhys.86.779} {\bibfield  {journal} {\bibinfo
  {journal} {Rev. Mod. Phys.}\ }\textbf {\bibinfo {volume} {86}},\ \bibinfo
  {pages} {779} (\bibinfo {year} {2014})}\BibitemShut {NoStop}%
\bibitem [{\citenamefont {Georges}\ \emph {et~al.}(1996)\citenamefont
  {Georges}, \citenamefont {Kotliar}, \citenamefont {Krauth},\ and\
  \citenamefont {Rozenberg}}]{georges1996dynamical}%
  \BibitemOpen
  \bibfield  {author} {\bibinfo {author} {\bibfnamefont {A.}~\bibnamefont
  {Georges}}, \bibinfo {author} {\bibfnamefont {G.}~\bibnamefont {Kotliar}},
  \bibinfo {author} {\bibfnamefont {W.}~\bibnamefont {Krauth}},\ and\ \bibinfo
  {author} {\bibfnamefont {M.~J.}\ \bibnamefont {Rozenberg}},\ }\href
  {https://doi.org/10.1103/RevModPhys.68.13} {\bibfield  {journal} {\bibinfo
  {journal} {Rev. Mod. Phys.}\ }\textbf {\bibinfo {volume} {68}},\ \bibinfo
  {pages} {13} (\bibinfo {year} {1996})}\BibitemShut {NoStop}%
\bibitem [{\citenamefont {Fotso}\ and\ \citenamefont
  {Freericks}(2020)}]{10.3389/fphy.2020.00324}%
  \BibitemOpen
  \bibfield  {author} {\bibinfo {author} {\bibfnamefont {H.~F.}\ \bibnamefont
  {Fotso}}\ and\ \bibinfo {author} {\bibfnamefont {J.~K.}\ \bibnamefont
  {Freericks}},\ }\bibfield  {journal} {\bibinfo  {journal} {Front. Phys.}\
  }\textbf {\bibinfo {volume} {8}},\ \href
  {https://doi.org/10.3389/fphy.2020.00324} {10.3389/fphy.2020.00324} (\bibinfo
  {year} {2020})\BibitemShut {NoStop}%
\bibitem [{\citenamefont {Yuen}\ \emph {et~al.}(2009)\citenamefont {Yuen},
  \citenamefont {Menon}, \citenamefont {Coates}, \citenamefont {Namdas},
  \citenamefont {Cho}, \citenamefont {Hannahs}, \citenamefont {Moses},\ and\
  \citenamefont {Heeger}}]{Yuen2009}%
  \BibitemOpen
  \bibfield  {author} {\bibinfo {author} {\bibfnamefont {J.~D.}\ \bibnamefont
  {Yuen}}, \bibinfo {author} {\bibfnamefont {R.}~\bibnamefont {Menon}},
  \bibinfo {author} {\bibfnamefont {N.~E.}\ \bibnamefont {Coates}}, \bibinfo
  {author} {\bibfnamefont {E.~B.}\ \bibnamefont {Namdas}}, \bibinfo {author}
  {\bibfnamefont {S.}~\bibnamefont {Cho}}, \bibinfo {author} {\bibfnamefont
  {S.~T.}\ \bibnamefont {Hannahs}}, \bibinfo {author} {\bibfnamefont
  {D.}~\bibnamefont {Moses}},\ and\ \bibinfo {author} {\bibfnamefont {A.~J.}\
  \bibnamefont {Heeger}},\ }\href {https://doi.org/10.1038/nmat2470} {\bibfield
   {journal} {\bibinfo  {journal} {Nat. Mat.}\ }\textbf {\bibinfo {volume}
  {8}},\ \bibinfo {pages} {572} (\bibinfo {year} {2009})}\BibitemShut {NoStop}%
\bibitem [{\citenamefont {Schwinger}\ \emph {et~al.}(2001)\citenamefont
  {Schwinger}, \citenamefont {Englert} \emph {et~al.}}]{schwinger2001quantum}%
  \BibitemOpen
  \bibfield  {author} {\bibinfo {author} {\bibfnamefont {J.}~\bibnamefont
  {Schwinger}}, \bibinfo {author} {\bibfnamefont {B.-G.}\ \bibnamefont
  {Englert}}, \emph {et~al.},\ }\href@noop {} {\emph {\bibinfo {title} {Quantum
  mechanics: symbolism of atomic measurements}}},\ Vol.~\bibinfo {volume} {1}\
  (\bibinfo  {publisher} {Springer},\ \bibinfo {year} {2001})\BibitemShut
  {NoStop}%
\bibitem [{\citenamefont {Han}\ \emph {et~al.}(2018)\citenamefont {Han},
  \citenamefont {Li}, \citenamefont {Aron},\ and\ \citenamefont
  {Kotliar}}]{hanPRB2018}%
  \BibitemOpen
  \bibfield  {author} {\bibinfo {author} {\bibfnamefont {J.~E.}\ \bibnamefont
  {Han}}, \bibinfo {author} {\bibfnamefont {J.}~\bibnamefont {Li}}, \bibinfo
  {author} {\bibfnamefont {C.}~\bibnamefont {Aron}},\ and\ \bibinfo {author}
  {\bibfnamefont {G.}~\bibnamefont {Kotliar}},\ }\href
  {https://doi.org/10.1103/PhysRevB.98.035145} {\bibfield  {journal} {\bibinfo
  {journal} {Phys. Rev. B}\ }\textbf {\bibinfo {volume} {98}},\ \bibinfo
  {pages} {035145} (\bibinfo {year} {2018})}\BibitemShut {NoStop}%
\bibitem [{\citenamefont {Antipov}\ \emph {et~al.}(2016)\citenamefont
  {Antipov}, \citenamefont {Javanmard}, \citenamefont {Ribeiro},\ and\
  \citenamefont {Kirchner}}]{PhysRevLett.117.146601}%
  \BibitemOpen
  \bibfield  {author} {\bibinfo {author} {\bibfnamefont {A.~E.}\ \bibnamefont
  {Antipov}}, \bibinfo {author} {\bibfnamefont {Y.}~\bibnamefont {Javanmard}},
  \bibinfo {author} {\bibfnamefont {P.}~\bibnamefont {Ribeiro}},\ and\ \bibinfo
  {author} {\bibfnamefont {S.}~\bibnamefont {Kirchner}},\ }\href
  {https://doi.org/10.1103/PhysRevLett.117.146601} {\bibfield  {journal}
  {\bibinfo  {journal} {Phys. Rev. Lett.}\ }\textbf {\bibinfo {volume} {117}},\
  \bibinfo {pages} {146601} (\bibinfo {year} {2016})}\BibitemShut {NoStop}%
\bibitem [{\citenamefont {Chen}\ and\ \citenamefont {Han}(2024)}]{xichen}%
  \BibitemOpen
  \bibfield  {author} {\bibinfo {author} {\bibfnamefont {X.}~\bibnamefont
  {Chen}}\ and\ \bibinfo {author} {\bibfnamefont {J.~E.}\ \bibnamefont {Han}},\
  }\href {https://doi.org/10.1103/PhysRevB.109.054307} {\bibfield  {journal}
  {\bibinfo  {journal} {Phys. Rev. B}\ }\textbf {\bibinfo {volume} {109}},\
  \bibinfo {pages} {054307} (\bibinfo {year} {2024})}\BibitemShut {NoStop}%
\bibitem [{\citenamefont {Soven}(1967)}]{PhysRev.156.809}%
  \BibitemOpen
  \bibfield  {author} {\bibinfo {author} {\bibfnamefont {P.}~\bibnamefont
  {Soven}},\ }\href {https://doi.org/10.1103/PhysRev.156.809} {\bibfield
  {journal} {\bibinfo  {journal} {Phys. Rev.}\ }\textbf {\bibinfo {volume}
  {156}},\ \bibinfo {pages} {809} (\bibinfo {year} {1967})}\BibitemShut
  {NoStop}%
\bibitem [{\citenamefont {Economou}(2006)}]{economou2006green}%
  \BibitemOpen
  \bibfield  {author} {\bibinfo {author} {\bibfnamefont {E.~N.}\ \bibnamefont
  {Economou}},\ }\href@noop {} {\emph {\bibinfo {title} {Green’s functions in
  quantum physics}}}\ (\bibinfo  {publisher} {Springer},\ \bibinfo {year}
  {2006})\BibitemShut {NoStop}%
\end{thebibliography}%

\end{document}